\documentclass[a4paper,12pt]{article}
\usepackage{jheppub} 
\usepackage{lineno}
\usepackage{subcaption}

\def\baa{\begin{array}}
\def\eaa{\end{array}}

\mathchardef\minus="002D

\def\({\left(}
\def\){\right)}
\def\[{\left[}
\def\]{\right]}

\def\hp{\hspace*{0.3mm}}

\def\hsm{\hspace*{-0.3mm}}

\title{\boldmath Curvature Perturbations from Higgs Modulated Reheating}

\author{Weiyi Deng$^{1}$,}
\emailAdd{dengwy23@mail2.sysu.edu.cn}

\author{Chengcheng Han$^{1,2}$,}
\emailAdd{hanchch@mail.sysu.edu.cn}

\author{Zhanhong Lei$^{1}$,}
\emailAdd{leizhh3@mail2.sysu.edu.cn}

\author{Jin Min Yang$^{3,4}$}
\emailAdd{jmyang@itp.ac.cn}

\vspace{0.5cm}

\affiliation{$^1$School of Physics, Sun Yat-Sen University, Guangzhou 510275, P. R. China}
\affiliation{$^2$Asia Pacific Center for Theoretical Physics, Pohang 37673, Republic of Korea}
\affiliation{$^3$ Center for Theoretical Physics, Henan Normal University, Xinxiang 453007,  P. R. China}
\affiliation{$^4$ Institute of Theoretical Physics, Chinese Academy of Sciences, Beijing 100190, P. R. China}
\vspace{0.5cm}

\abstract{In this work we investigate curvature perturbations and non-Gaussianity arising from Higgs modulated reheating in the early Universe. We employ three different methods — the period-averaging (PA) method, the exact method, and the non-perturbative $\delta N$ formalism — to compute the power spectrum and bispectrum of curvature perturbations. Our results show that the non-perturbative $\delta N$ method provides a reliable estimate across a wide range of reheating time and Higgs field values, including regimes where the Higgs field oscillates significantly after inflation. We find that a smaller Higgs self-coupling ($\lambda$) leads to a larger curvature perturbation, with the non-Gaussianity predominantly taking a local shape. This highlights the importance of considering non-perturbative effects in calculating the curvature perturbation during Higgs modulated reheating, especially for smaller values of $\lambda$. Our findings offer valuable insights into the dynamics of reheating and the generation of primordial perturbations in the early Universe. }

\begin{document}
\maketitle
\flushbottom

\section{Introduction}
\label{sec:intro}

It is widely believed that the early Universe underwent a period of rapid expansion, known as inflation. During this stage, the Hubble parameter remained nearly constant, and the Universe was well approximated by a de Sitter spacetime. The origin of this phase is typically attributed to the dynamics of the inflaton field. Quantum fluctuations of the inflaton are thought to source the primordial curvature perturbations~\cite{Guth:1982ec,Hawking:1982cz,Bardeen:1983qw,Lyth:1984gv,Mukhanov:1981xt,Starobinsky:1982ee}, which eventually seed the large-scale structure of the Universe.
However, it has also been shown that spatial variations in the inflaton decay rate $\Gamma$ at the end of inflation can generate primordial curvature perturbations~\cite{Dvali:2003em,Kofman:2003nx,Suyama:2007bg,Ichikawa:2008ne}. This mechanism is known as the \emph{modulated reheating scenario}. In such a setup, it is often assumed that a light, weakly interacting scalar field acquires a large vacuum expectation value (VEV) during inflation. This VEV varies across different Hubble patches due to quantum fluctuations of the field. If the inflaton’s interactions with other particles are  modulated by the fluctuations of this light scalar field,  the inflaton decay rate varies across different regions of the universe, leading to spatially dependent reheating.

However, during inflation an interacting scalar field---such as a Higgs-like particle---can also develop a VEV that varies across different patches of the Universe. 
In this case, modulated reheating can be triggered by a Higgs field with self-interactions. 
The Higgs-like field $h$ acquires quantum fluctuations during inflation, whose probability distribution can be described by the Fokker–Planck equation~\cite{Starobinsky:1986fx,Starobinsky:1994bd,Markkanen:2019kpv},
\begin{equation}
\frac{\partial \rho(h,t)}{\partial t}
= \frac{1}{3H_{\mathrm{inf}}}\frac{\partial}{\partial h}
\left[V'(h)\rho(h,t)\right]
+ \frac{H_{\mathrm{inf}}^3}{8\pi^2}\frac{\partial^2\rho(h,t)}{\partial h^2}~,
\label{eq:FP}
\end{equation}
where $V(h)=\frac{\lambda}{4}h^4$ is the Higgs potential, $H_{\mathrm{inf}}$ is the Hubble parameter during inflation, and $\rho(h,t)$ is the probability distribution of the Higgs-like field $h$ across the Universe. 
For a sufficiently long inflationary period, a Higgs-like scalar with self-coupling $\lambda$ reaches an equilibrium distribution,
\begin{equation}
\rho_{\mathrm{eq}}(h)
=\left(\frac{32\pi^2\lambda}{3}\right)^{1/4}
\frac{1}{\Gamma\!\left(\frac14\right)H_{\mathrm{inf}}}
\exp\!\left(-\frac{2\pi^2\lambda h^4}{3H_{\mathrm{inf}}^4}\right)~,
\label{eq:rhoeq}
\end{equation}
with a typical VEV of order $\mathcal{O}(H_{\mathrm{inf}})$ varying across different Hubble patches. 
This opens up the possibility that the Higgs field itself could act as the modulating field for the inflaton decay.

While Higgs modulated reheating has been studied in the literature~\cite{Choi:2012cp,DeSimone:2012gq,Cai:2013caa,Karam:2021qgn,Litsa:2020mvj,Litsa:2020rsm,Lu:2019tjj,You:2024hit,Han:2024qbw}, most analyses assume that the Higgs value remains constant after inflation. 
However, unlike a light non-interacting scalar field, the Higgs field continues to evolve after inflation. 
As pointed out in~\cite{Lu:2019tjj,You:2024hit,Han:2024qbw}, for $t \lesssim t_c=\frac{\sqrt{2}}{3\sqrt{\lambda}h_{\mathrm{inf}}}$, the Higgs field value remains nearly constant, but for $t \gtrsim t_c$, it evolves as
\begin{equation}
h(t) \propto t^{-2/3}\,|h_{\mathrm{inf}}|^{1/3}
\cos\!\left(\omega|h_{\mathrm{inf}}|^{1/3} + \theta\right)~,
\label{eq:hreh}
\end{equation}
where $\omega=2.3\lambda^{\frac{1}{6}}t^{\frac{1}{3}}$, and $h_\mathrm{inf}$ denotes the Higgs field value at the end of inflation. This 
indicates that the Higgs field not only decreases but also oscillates after inflation. 
Consequently, the relation between the curvature perturbation and the Higgs field becomes highly non-linear. 
If one continues to use the $\delta N$ formalism~\cite{Starobinsky:1985ibc,Salopek:1990jq,Comer:1994np,Sasaki:1995aw,Sasaki:1998ug,Wands:2000dp,Lyth:2003im,Rigopoulos:2003ak,Lyth:2004gb,Lyth:2005fi}, a subtle issue arises: when $\omega$ is large enough, the $\delta N$ expansion becomes invalid because each derivative of $N$ introduces a factor of $\omega$, so that higher-order terms are not guaranteed to be smaller than lower-order ones. 
Then a key question is how to properly treat such highly non-linear effects in calculating the $n$-point correlation functions for the curvature perturbation.

In this work, we adopt several different methods to calculate the two-point and three-point correlation functions of the curvature perturbations and compare them systematically. 
We focus on the weakly interacting regime $\lambda \lesssim \mathcal{O}(10^{-3})$, in which the Higgs field can be treated as nearly massless. 
Therefore in our observable Universe, the Higgs field could be approximated by a Gaussian distribution centered around some average value $\bar{h}$. 
Since $\bar{h}$ varies across the entire Universe, our local patch is not necessarily representative of the typical value. 
Although $\bar{h}$ cannot be uniquely determined, we illustrate our results for representative choices of $\bar{h}$.

The paper is organized as follows. 
In Sec.~\ref{sec:higgs_modul_reh} we briefly review how primordial curvature perturbations are generated through Higgs-modulated inflaton decay during reheating. 
In Sec.~\ref{sec:method} we present several computational approaches, including the period-averaging method, the exact method and the non-perturbative $\delta N$ formalism. 
Numerical results are presented in Sec.~\ref{sec:numerical}, where we map out the validity of each method. 
In particular, we show that the naive $\delta N$ expansion applies when reheating occurs early, while the period-averaging method is suitable for reheating occurring much later. 
The non-perturbative $\delta N$ formalism performs well across all regimes, including the intermediate stage where oscillations are significant. 
We summarize our findings in Sec.~\ref{sec:conclusion}.

\section{Higgs Modulated Reheating}
\label{sec:higgs_modul_reh}

In this section, we present the essential ingredients of Higgs modulated reheating. To realize this mechanism, we assume that the inflaton decay rate $\Gamma$ depends on $h_{\mathrm{reh}}$, the Higgs field value at the time of reheating. Taking an example, here we assume the function of $\Gamma(h_{\mathrm{reh}})$ can be approximated as
\begin{equation}
\Gamma(h_{\mathrm{reh}})= \Gamma_0 \left(1+\frac{h^2_{\mathrm{reh}}}{\Lambda^2} \right)~,
\end{equation}
where $\Gamma_0$ represents the constant part that dominates the inflaton decay, while the total decay rate is modulated by the Higgs field. Notice that $h_{\mathrm{reh}}$ generally differs from the Higgs field value immediately after inflation due to the post-inflationary evolution of the field. The parameter $\Lambda$ characterizes the strength of Higgs modulation and is determined by the underlying UV model. In this work, we assume $\Lambda \gg h_{\mathrm{reh}}$ so that the inflaton decay rate is still primarily governed by  $\Gamma_0$, with the modulation providing only a small perturbative correction. This form of Higgs modulated reheating can naturally arise, for example, in scenarios where the inflaton couples to right-handed neutrinos~\cite{You:2024hit,Han:2024qbw}. We emphasize that our method for computing the curvature perturbation does not rely on this specific functional form of $\Gamma(h_{\mathrm{reh}})$.

During inflation, the Higgs field can be decomposed as
\begin{equation}
h=\bar{h} + \delta h~,
\end{equation}
where $\bar{h}$	is the average Higgs value in our observable Universe and 
$\delta h$ represents its fluctuation. Neglecting the Higgs self-interaction, $\delta h$ can be treated as a Gaussian random field. The two-point correlation function for a massless Gaussian field is given by 
\begin{equation}
\langle \delta h_{\bf k} \delta h_{\bf k^\prime}\rangle = (2\pi)^3 \delta^3({\bf k-k^\prime})P_{\delta h}(k),
\end{equation}
where 
\begin{equation}
P_{\delta h}(k)=\frac{H_{\mathrm{inf}}^2}{2k^3}~,
\end{equation}
with $H_{\mathrm{inf}}$ being the Hubble parameter during inflation.

After inflation, we assume the inflaton oscillates around the minimum of a quadratic potential. Averaging over its oscillation period, the Universe undergoes a matter-dominated–like expansion, characterized by
\begin{equation}
a(t)\sim t^{2/3}, ~~~H(t)=\frac{2}{3t}~.
\end{equation}
The post-inflationary evolution of the Higgs field is governed by the Klein–Gordon equation,
\begin{equation}
\label{eq:KG_higgs}
\ddot{h}(t)+\frac{2}{t}\dot{h}(t)+\lambda h^3(t)=0~,
\end{equation}
with the initial condition $h_{\mathrm{inf}}=h(\mathbf{x},t_{\mathrm{end}})$ where $h(\mathbf{x},t_{\mathrm{end}})$ is the Higgs value at the end of inflation.

At the beginning of the inflaton oscillation stage, the Higgs field remains nearly constant for a time
\begin{equation}
t_{\mathrm{c}}= \frac{\sqrt{2}}{3\sqrt{\lambda}h_{\mathrm{inf}}}~.
\end{equation}
For $t> t_c$, the Higgs field starts to decrease and oscillate. A semi-analytical solution to Eq.~\eqref{eq:KG_higgs} can be derived \cite{You:2024hit,Han:2024qbw}:
\begin{equation}
h(t)\simeq \mathrm{sgn}(h_{\mathrm{inf}}) A\left(\frac{|h_{\mathrm{inf}}|}{\lambda}\right)^{\frac{1}{3}} t^{-2/3}\cos\left(\alpha \lambda^{\frac{1}{6}}|h_{\mathrm{inf}}|^{\frac{1}{3}}  t^{\frac{1}{3}}+\theta\right)~,
\end{equation}
where $A\simeq0.9$, $\alpha= \frac{\,\Gamma^2(3/4)\,}{\sqrt{\pi\,}}
{6}^{\frac{1}{3}} 5^{\frac 1 4}=2.3$, and $\theta= -3^{-1/3}2^{1/6}\alpha-\arctan 2 =  -2.9$. At the time when reheating completed, $t=t_\mathrm{reh}$, the Higgs field takes the form of an oscillating function of $h_{\mathrm{inf}}$,
\begin{equation}
h_{\mathrm{reh}}= h(t_\mathrm{reh},h_{\mathrm{inf}})\propto |h_{\mathrm{inf}}|^{\frac{1}{3}}\cos(\omega|h_{\mathrm{inf}}|^{\frac{1}{3}}+\theta),\quad\omega=\lambda^{\frac{1}{6}}t_{\mathrm{reh}}^{\frac{1}{3}}\alpha~.
\end{equation}

The curvature perturbation generated from Higgs modulated reheating, $\zeta_h(\mathbf{x},t)$, can be calculated using the $\delta N$ formalism~\cite{Starobinsky:1985ibc,Salopek:1990jq,Comer:1994np,Sasaki:1995aw,Sasaki:1998ug,Wands:2000dp,Lyth:2003im,Rigopoulos:2003ak,Lyth:2004gb,Lyth:2005fi}, 
\begin{equation}
	\zeta_h(\mathbf{x},t)=\delta N(\mathbf{x},t)=N(\mathbf{x},t)-\langle N(\mathbf{x},t)\rangle~,
\end{equation}
where $N(\mathbf{x},t)$ is the number of e-folds of the cosmic expansion from the time inflation ends to some time $t$ after reheating under the uniform energy density gauge,
\begin{equation}
	\begin{aligned}
	N(\mathbf{x},t)&=\int_{t_{\mathrm{end}}}^{t_{\mathrm{reh}}(\mathbf{x})}\mathrm{d}t^\prime H(t^\prime)+\int_{t_{\mathrm{reh}}(\mathbf{x})}^{t}\mathrm{d}t^\prime H(t^\prime)\\&=\int_{\rho_{\mathrm{inf}}}^{\rho_{\mathrm{reh}}(h(\mathbf{x}))}\mathrm{d}\rho\frac{H}{\dot{\rho}}+\int_{\rho_{\mathrm{reh}}(h(\mathbf{x}))}^{\rho(t)}\mathrm{d}\rho\frac{H}{\dot{\rho}}~,
	\end{aligned}
\end{equation}
with $\rho$ being the total energy density of the Universe, $\rho_{\mathrm{inf}}=\rho(t_{\mathrm{end}})$ and $\rho_{\mathrm{reh}}=\rho(t_{\mathrm{reh}})$. Utilizing the continuity equation
\begin{equation}
	\dot{\rho}+3H(1+w)\rho=0,
\end{equation}
where $\omega=0$ before $t_{\mathrm{reh}}$ and $\omega=1/3$ after $t_{\mathrm{reh}}$, we obtain
\begin{equation}
N(\mathbf{x},t)=-\frac{1}{3}\ln\frac{\rho_{\mathrm{reh}}\left(h(\mathbf{x})\right)}{\rho_{\mathrm{inf}}}-\frac{1}{4}\ln\frac{\rho(t)}{\rho_{\mathrm{reh}}\left(h(\mathbf{x})\right)}~.
\end{equation}
Then $\zeta_h(\mathrm{x},t>t_{\mathrm{reh}})$ can be derived as
\begin{equation}
\begin{aligned}\zeta_{h}(\mathbf{x},t>t_{\mathrm{reh}})=\delta N
&=N(\mathbf{x},t)-\langle N(\mathbf{x},t)\rangle\\
&=-\frac{1}{12}\left[\ln\rho_{\mathrm{reh}}(\mathbf{x})-\langle\ln\rho_{\mathrm{reh}}(\mathbf{x})\rangle\right]\\&=-\frac{1}{6}\left[\ln(\Gamma)-\langle\ln(\Gamma)\rangle\right]~,
\end{aligned}
\end{equation}
where we have used the Friedmann equation $\rho=3H^2M_p^2$ and the relation $H(t_{\mathrm{reh}})=\Gamma$. We observe that the comoving curvature perturbation $\zeta_{h}$ is generated by the modulated decay rate $\Gamma(h(t_{\mathrm{reh}},h_{\mathrm{inf}}))$, which is in turn determined by the fluctuation of the Higgs field value at the end of inflation across different Hubble patches, $h_{\mathrm{inf}}(\mathbf{x})=h(\mathbf{x},t_{\mathrm{end}})$.

The curvature perturbation $\zeta_h$ is ultimately a function of $h_{\mathrm{reh}}$, and therefore of $h_{\mathrm{inf}}$. Note that the highly non-linear relation between the $h_{\mathrm{reh}}$ and $h_{\mathrm{inf}}$,
\begin{equation}
h_{\mathrm{reh}}\propto |h_{\mathrm{inf}}|^{\frac{1}{3}}\cos(\omega|h_{\mathrm{inf}}|^{\frac{1}{3}}+\theta)~,
\end{equation}
poses difficulties for the standard $\delta N$ expansion
\begin{equation}
\zeta_h= \delta N = N' \delta h_{\mathrm{inf}} +\frac{1}{2} N'' \delta h_{\mathrm{inf}}^2+...
\end{equation}
As pointed in Refs.\cite{You:2024hit,Han:2024qbw}, the $\delta N$ expansion becomes invalid when $\omega$ becomes large. The reason is that each derivative would get a factor of $\omega$, thus the higher-order terms are not guaranteed to be smaller than lower-order ones. To circumvent this issue, Refs.~\cite{You:2024hit,Han:2024qbw} introduced a period-averaging (PA) method, in which the square of cosine term is averaged to $1/2$ in the large-$\omega$ limit. In this work, we also employ an all-order treatment together with a non-perturbative $\delta N$ method to evaluate the curvature perturbation. We then compare the resulting two-point and three-point correlation functions of the curvature perturbation across these different approaches.

\section{Methods for curvature perturbation calculation}
\label{sec:method}
In this section we briefly overview three main methods to calculate the curvature perturbation: the period-averaging(PA) method, the exact method and the non-perturbative $\delta N$ method.
\subsection{Period-averaging(PA) method}

We first adopt the period-averaging method to calculate the curvature perturbation. Since $\zeta_h$ is a function of $h^2_{\mathrm{reh}}$, this induce a factor square of cosine when translating into the function of $h_{\mathrm{inf}}$. When $\omega$ is large enough, we can take square of cosine as $1/2$ for a period, then the period-averaged curvature perturbation is no longer oscillatory and we can safely do the $\delta N$ expansion, which we call the period-averaging method. Since the expansion is always conducted at the spatially-flat hypersurface at the end of inflation, in the following we denote $h_\mathrm{inf}(\mathbf{x})$ as $h(\mathbf{x})$ and $\delta h_\mathrm{inf}$ as $\delta h$ for notational convenience, therefore at $t_\mathrm{end}$ the field perturbation can be expanded as $h(\mathbf{x})=\bar{h}+\delta h(\mathbf{x})$. Now the period-averaged $\zeta_{h}$ can be expanded in terms of $\delta h$ around $\bar{h}$,
\begin{equation}
\zeta_{h}\simeq z_{1}\delta h(\mathbf{x})+\frac{1}{2}z_{2}\delta h^{2}(\mathbf{x})~.
\end{equation}
The Taylor expansion coefficients $z_1$ and $z_2$ can be written as
\begin{equation}
z_{1}=-\frac{1}{6}\left.\frac{\Gamma^{\prime}}{\Gamma}\right|_{\bar{h}},\quad z_{2}=-\frac{1}{6}\left.\left[\frac{\Gamma^{\prime\prime}}{\Gamma}-\left(\frac{\Gamma^{\prime}}{\Gamma}\right)^{2}\right]\right|_{\bar{h}}~,
\end{equation}
where $\Gamma$ is the period-averaged decay rate.

To the leading order, the two-point correlation function of $\zeta_h$ is given by
\begin{equation}
\langle\zeta_h(\mathbf{k})\zeta_h(\mathbf{k}^\prime)\rangle^{\text{PA}}=z_1^2\langle\delta h(\mathbf{k})\delta h(\mathbf{k}^\prime)\rangle~.
\end{equation}
Therefore the power spectrum of $\zeta_h$ is
\begin{equation}
\label{eq:power_PA}
\mathcal{P}_{\zeta}^{(h)}=z_{1}^{2}\mathcal{P}_{\delta h}=\frac{z_{1}^{2}H_{\mathrm{inf}}^{2}}{4\pi^{2}}~,
\end{equation}
where $\mathcal{P}_{\delta h}$ is the dimensionless power spectrum defined as $\mathcal{P}_{\delta h}=\frac{k^3}{2\pi^2}P_{\delta h}(k)=\frac{H_{\mathrm{inf}}^{2}}{4\pi^2}$.

The total comoving curvature perturbation after reheating comprises contributions from both the inflaton perturbation and the Higgs modulated reheating, $\zeta=\zeta_\phi+\zeta_h$, which are uncorrelated. Thus the power spectrum of $\zeta$ is given by
\begin{equation}
\mathcal{P}_{\zeta}=\mathcal{P}_{\zeta}^{(\phi)}+\mathcal{P}_{\zeta}^{(h)}~,
\end{equation}
with the measured amplitude $\mathcal{P}_{\zeta}^{(o)}\simeq2.1\times10^{-9}$ at the pivot scale $k_*=0.05\,\mathrm{Mpc}^{-1}$ from the Planck data~\cite{Planck:2018nkj,Planck:2018jri}. Define $R$ as the square root of the ratio between the power spectra of $\zeta_{h}$ and the observed $\zeta$,
\begin{equation}
R\equiv\left(\frac{\mathcal{P}_{\zeta}^{(h)}}{\mathcal{P}_{\zeta}^{(o)}}\right)^{1/2}~.
\end{equation}
To agree with the observation, it is required that $R<1$.

The calculation of the three-point correlation function of the curvature perturbation is similar. To the leading order, we obtain
\begin{equation}
\begin{aligned}
\langle\zeta_h(\mathbf{k}_{1})\zeta_h(\mathbf{k}_{2})\zeta_h(\mathbf{k}_{3})\rangle&=(2\pi)^{3}\delta^3(\mathbf{k_1}+\mathbf{k_2}+\mathbf{k_3})B^{(h)}_\zeta(k_1,k_2,k_3)\\
&=(2\pi)^{3}\delta^3(\mathbf{k_1}+\mathbf{k_2}+\mathbf{k_3})z_1^2z_2[P_{\delta h}(k_{1})P_{\delta h}(k_{2})+\mathrm{cyclic}]~.
\end{aligned}
\end{equation}
Although the field fluctuation is assumed to be Gaussian, the nonlinear relation between $\zeta_{h}$ and $\delta h$ gives rise to a local-shape bispectrum. 

The local-shape bispectrum is conventionally parameterized by a single parameter $f^{\mathrm{local}}_{\mathrm{NL}}$ defined as~\cite{Komatsu:2001rj,Maldacena:2002vr,Lyth:2005fi,Wands:2010af}
\begin{equation}
	B^{(h)}_\zeta(k_1,k_2,k_3)=\frac{6}{5}f^{\mathrm{local}}_{\mathrm{NL}}[P_\zeta(k_1)P_\zeta(k_2)+\mathrm{cyclic}]~,
\end{equation}
therefore the bispectrum generated from Higgs modulated reheating can be parameterized as
\begin{equation}
\label{eq:fnl_PA}
	f^{\mathrm{local}}_{\mathrm{NL}}=\frac{5}{6}\frac{z_2}{z_1^2}R^4~.
\end{equation}
However, according to its definition, the period-averaging approximation works well in the large-$\omega$ limit, as shown below.

\subsection{Exact method}

To obtain results free from approximations and valid at all orders, we go back to the definition of the $x$-space correlation function for curvature perturbation. The exact definition of the $m$-point function for $\zeta_h$ is given by
\begin{equation}
\label{eq:exact_cor_func}
\begin{aligned}
\langle\zeta_{1}\cdots\zeta_{m}\rangle&=\langle(N_{1}-\bar{N})\cdots(N_{m}-\bar{N})\rangle\\&=\int\mathrm{d}h_{1}\cdots\int\mathrm{d}h_{m}(N_{1}-\bar{N})\cdots(N_{m}-\bar{N})p(h_1,\cdots,h_m)~,
\end{aligned}	
\end{equation}
where the subscript $i$  denotes the spatial position, for example, $\zeta_i=\zeta_h(\mathbf{x}_i)$, $N_i=N(\mathbf{x}_i)$, and $p(h_1,\cdots,h_m)$ is the joint probability distribution function of the $m$ fields $h_i=h(\mathbf{x}_i)$. If $N$ is a simple function of $h$ and $p(h_1,\cdots,h_m)$ is Gaussian distribution, then Eq.~\eqref{eq:exact_cor_func} can be evaluated analytically. For more general cases numerical computation becomes necessary. 

Neglecting the Higgs self-interaction, $p(h_1,\cdots,h_m)$ becomes a $m$-dimensional Gaussian distribution with mean value $\bar{h}$ and covariance matrix $\Sigma_{ij}=\Sigma(r_{ij})=\langle\delta h(\mathbf{x}_i)\delta h(\mathbf{x}_j)\rangle$. The Gaussian distribution $p_G$ is given by 
\begin{equation}
\label{eq:Gauss_mpdf}
    p_G(h_1,\cdots,h_m)=(2\pi)^{-m/2}|\Sigma|^{-1/2}\exp{\left(-\frac{1}{2}\delta h_i(\Sigma^{-1})_{ij}\delta h_j \right)}~,
\end{equation}
where $\delta h_i=h(\mathbf{x}_i)-\bar{h}$. Since $\delta h$ is a Gaussian perturbation with scale-invariant power spectrum, one can obtain
\begin{equation}
\label{eq:sigma_ij}
\begin{aligned}
\Sigma(r_{ij})&=\langle\delta h(\mathbf{x}_i)\delta h(\mathbf{x}_j)\rangle=\int\frac{d^3k_1}{(2\pi)^3}\frac{d^3k_2}{(2\pi)^3}\langle\delta h_{\mathbf{k}_1}\delta h_{\mathbf{k}_2}\rangle e^{i(\mathbf{k}_1\mathbf{x}_i+\mathbf{k}_2\mathbf{x}_j)}\\
&=\int\frac{d^3k}{(2\pi)^3}P_{\delta h}(k) e^{i\mathbf{k}(\mathbf{x}_i-\mathbf{x}_j)}=\frac{H^2_{\mathrm{inf}}}{4\pi^2}\int^{k_{\mathrm{max}}}_{k_{\min}}dk\frac{\sin kr_{ij}}{k^2r_{ij}}~,
\end{aligned}
\end{equation}  
where we have introduced a UV cutoff $k_{\mathrm{max}}$ and an IR cutoff $k_{\mathrm{min}}$. Here $r_{ij}=|\mathbf{x}_i-\mathbf{x}_j|$. In this case, the IR cutoff is taken to be the size of our observable Universe, $k_{\mathrm{min}}=\frac{2\pi}{L_{\mathrm{obs}}}$, with $L_\mathrm{obs}\sim 14~\mathrm{Gpc}$. The UV cutoff corresponds to the modes exiting the horizon right before the end of inflation, i.e. $k_{\mathrm{max}}\sim e^{N_e} k_{\mathrm{min}}$, where $N_e\sim60$ is the number of e-folds between horizon-crossing and the end of inflation. Integrating Eq.~\eqref{eq:sigma_ij} gives
\small
\begin{equation}
\Sigma(r)\equiv \langle\delta h(0)\delta h(\mathbf{x})\rangle=\frac{H^2_{\mathrm{inf}}}{4\pi^2}\left[\mathcal{C}_i(k_{\mathrm{max}}r)-\mathcal{C}_i(k_{\mathrm{min}}r)+\frac{\sin (k_{\mathrm{min}}r)}{k_{\mathrm{min}}r}-\frac{\sin (k_{\mathrm{max}}r)}{k_{\mathrm{max}}r}\right]~,
\end{equation}
\normalsize
where $\mathcal{C}_i$ is the cosine integral function
\begin{equation}
	\mathcal{C}_i(x)=-\int_x^\infty\frac{\cos(t)}{t}dt~.
\end{equation}
Using Eq.~\eqref{eq:exact_cor_func}, we could get the two point correlation function $\langle\zeta_h(\mathbf{x}_1)\zeta_h(\mathbf{x}_2)\rangle$. Then the power spectrum can be obtained by performing
a Fourier transforma on the two-point correlation function,
\begin{eqnarray}
\label{eq:exact_power}
    P^{(h)}_{\zeta}(k)&=&\int_{-\infty}^{\infty}\mathrm{d}^{3}\mathbf{r_{12}}\langle\zeta_h(\mathbf{x}_1)\zeta_h(\mathbf{x}_2)\rangle e^{-i\mathbf{k}\cdot\mathbf{r_{12}}} \nonumber \\
    &=&4\pi\int_0^\infty\mathrm{d}r_{12}r_{12}^2\langle\zeta_h(\mathbf{x}_1)\zeta_h(\mathbf{x}_2)\rangle\frac{\sin(kr_{12})}{kr_{12}}~,
\end{eqnarray}
where $\mathbf{r}_{12}=\mathbf{x}_1-\mathbf{x}_2$, and $\langle\zeta_h(\mathbf{x}_1)\zeta_h(\mathbf{x}_2)\rangle$ is a function of $r_{12}$ due to the invariance of translation and rotation. However, the integrand at the right hand side of Eq.~\eqref{eq:exact_power} is highly oscillatory. In order to obtain accurate results, we employ the Fast Fourier Transform (FFT) algorithm to compute the power spectrum. The dimensionless power spectrum is obtained by
\begin{equation}
\label{eq:power_exact}
    \mathcal{P}_\zeta^{(h)}(k)=\frac{k^3}{2\pi^2}P_\zeta^{(h)\mathrm{FFT}}(k)~,
\end{equation}
where $P_\zeta^{(h)\mathrm{FFT}}(k)$ is calculated using the FFT algorithm. 

The three-point correlation function in $x$-space can be directly computed by treating $p(h_1,\dots,h_m)$ as a three-dimensional Gaussian distribution with mean $\bar{h}$ and covariance matrix $\Sigma_{ij} = \Sigma(r_{ij}) = \langle \delta h(\mathbf{x}_i),\delta h(\mathbf{x}_j)\rangle$. The bispectrum in $k$-space can, in principle, be obtained via Fourier transformation of the three-point correlation function in $x$-space. However, this requires a six-dimensional Fourier transform, which is numerically very challenging. Instead, we compare the three-point functions in $x$-space obtained with other methods.

\subsection{Non-perturbative $\delta N$ method}

In general, the calculation of Eq.~\eqref{eq:exact_cor_func} is rather cumbersome, whereas the period-averaging method may provide a valid approximation for sufficiently large values of $\omega$. The non-perturbative $\delta N$ approach proposed in Refs.~\cite{Suyama:2013dqa, Imrith:2018uyk} offers an alternative approximation scheme that is particularly well suited to scenarios where $N$ exhibits oscillatory behavior. In the following we give a brief introduction of this method, with full details provided in Appendix. \ref{app:non_perturb}. Assuming negligible Higgs self-interaction, the joint probability distribution $p(h_1,\cdots,h_m)$ can be approximated by the $m$-dimensional Gaussian distribution $p_G(h_1,\cdots,h_m)$. We then define
\begin{equation}
\xi_{ij} = \frac{\Sigma_{ij}}{\Sigma(0)}~,
\end{equation}
which satisfies $\xi_{ij} < 1$. For the scales of interest that can be measured by CMB observations, $\xi$ remains much smaller than unity; for instance, $\xi(r_*)=\xi(0.01L_\mathrm{obs})\simeq 0.05$. In this regime, $p_G(h_1,\cdots,h_m)$ can be Taylor expanded in terms of $\xi_{ij}$, allowing Eq.~\eqref{eq:exact_cor_func} for the exact correlation function to be considerably simplified.


For illustration, we first consider the two-point function. From the definition of $m$-dimensional Gaussian distribution in Eq.~\eqref{eq:Gauss_mpdf}, we can derive the following relation using matrix calculus,
\begin{equation}
    \frac{\partial p_G}{\partial\Sigma_{ij}}=p_G\left[-\frac{1}{2}\Sigma^{-1}_{ij}+\frac{1}{2}\delta h_k\delta h_l\Sigma^{-1}_{ik}\Sigma^{-1}_{jl} \right]~.
\end{equation}
Considering $\xi_{12}\ll1$, the two-point distribution $p_G(h_1,h_2)$ with $\Sigma_{ij}=\Sigma(0)\left(\begin{array}{cc}1&\xi_{12} \\ \xi_{12} & 1 \end{array}\right)$ and $\Sigma^{-1}_{ij}\simeq\Sigma(0)^{-1}\left(\begin{array}{cc}1&-\xi_{12} \\ -\xi_{12} & 1 \end{array}\right)$ can be expanded around $\xi_{12}$ as
\begin{equation}
\begin{aligned}
p_G(h_1,h_2)&=\left. p_G(h_1,h_2)\right|_{\xi_{12}=0}+2\Sigma(0)  \left.\frac{\partial p_G(h_1,h_2)}{\partial\Sigma_{12}}\right|_{\xi_{12}=0} \xi_{12}+\mathcal{O}(\xi_{12}^2)\\    &=p_G(h_1)p_G(h_2)+p_G(h_1)p_G(h_2)\delta h_1\delta h_2\Sigma(0)^{-1} \xi_{12} + \mathcal{O}(\xi_{12}^2)~.
\end{aligned}
\end{equation}
Substituting this expansion into $\langle\zeta_1\zeta_2\rangle$, we obtain 
\begin{equation}
\begin{aligned}
\langle\zeta_1\zeta_2\rangle&=\int \mathrm{d}h_1\mathrm{d}h_2(N_1-\bar{N})(N_2-\bar{N})p_G(h_1,h_2)\\
&\simeq \left[\int\mathrm{d}h(N-\bar{N})p_G(h)\right]^2+\Sigma(0)^{-1}\xi_{12}\left[\int\mathrm{d}h(N-\bar{N})p_G(h)\delta h\right]^2\\
&=\Sigma_{12}\Sigma(0)^{-2}\left[\int\mathrm{d}hNp_G(h)\delta h\right]^2~.
\end{aligned}
\end{equation}
Therefore, if we define the so-called non-perturbative $\delta N$ coefficient $\tilde{N}^\prime$ as 
\begin{equation}
    \tilde{N}^\prime=\Sigma^{-1}(0)\int\mathrm{d}h ~p_\mathrm{G}(h)N\delta h~,
\end{equation}
the two-point function then takes a very simple form,
\begin{equation}    \langle\zeta_h(\mathbf{x}_1)\zeta_h(\mathbf{x}_2)\rangle\approx\tilde{N}^{\prime2}\Sigma(r_{12})~.
\end{equation}
Formally, this expression is similar to the leading order expansion of the standard $\delta N$ formalism, with the coefficient $N^\prime$ replaced by a statistical average coefficient $\tilde{N}^\prime$. When $\zeta$ is a slowly-varying function of the field $h$, $\tilde{N}^\prime$ simply reduces to $N^\prime$. However, if $\zeta$ is a rapidly oscillating function of $h$, $\tilde{N}^\prime$ serves as a non-perturbative extension of $N^\prime$ and remains valid when the naive $\delta N$ expansion breaks down. 

Following similar procedures, the three-point function of the curvature perturbation can be simplified as
\begin{equation}
\label{eq:3pt_NP}
\langle\zeta_h(\mathbf{x}_1)\zeta_h(\mathbf{x}_2)\zeta_h(\mathbf{x}_3)\rangle\approx\tilde{N}^{\prime\prime}\tilde{N}^{\prime2}\left[\Sigma(r_{12})\Sigma(r_{23})+\mathrm{cyclic}\right]~,
\end{equation}
where $\tilde{N}^{\prime\prime}$ is another non-perturbative $\delta N$ coefficient defined as
\begin{equation}
\tilde{N}^{\prime\prime}=\Sigma^{-2}(0)\int\mathrm{d}h ~p_{\mathrm{G}}(h)(N-\bar{N})\delta h^2~.
\end{equation}

Apart from greatly simplifying the calculation of correlation functions, another advantage of the non-perturbative $\delta N$ method is that the power spectrum and bispectrum of curvature perturbations in $k$-space can be expressed quite simply in terms of non-perturbative $\delta N$ coefficients,
\begin{equation}
\label{eq:spectrum_NP}
\begin{aligned}
P^{(h)}_{\zeta}(k)&\approx\tilde{N}^{\prime2}P_{\delta h}(k)~,\\
B^{(h)}_\zeta(k_1,k_2,k_3)&\approx\tilde{N}^{\prime2}\tilde{N}^{\prime\prime}(P_{\delta h}(k_{1})P_{\delta h}(k_{2})+\mathrm{cyclic}).
\end{aligned}
\end{equation}
Similarly, the dimensionless power spectrum and the reduced bispectrum are given by
\begin{equation}
\label{eq:reduced_spectrum_NP}
\begin{aligned}
\mathcal{P}_{\zeta}^{(h)}&=\tilde{N}^{\prime2}\mathcal{P}_{\delta h}=\frac{\tilde{N}^{\prime2}H_{\mathrm{inf}}^{2}}{4\pi^{2}},\\
f^{\mathrm{local}}_{\mathrm{NL}}&=\frac{5}{6}\frac{\tilde{N}^{\prime\prime}}{\tilde{N}^{\prime2}}R^4~.
\end{aligned}
\end{equation}

\section{Numerical results}
\label{sec:numerical}
In this section we provide our numerical results for different methods. To present our result, we fix $\Lambda=10H_{\mathrm{inf}}$ and $\lambda=0.001$. The Higgs value at $t_{\mathrm{reh}}$ is given by
\begin{equation}
h_{\mathrm{reh}}\simeq\left\{
\begin{aligned}
&h,&t \leq  t_c~, \\
&A_h |h|^{\frac{1}{3}}\cos(\omega|h|^{\frac{1}{3}}+\theta), &t>t_c~,
\end{aligned}\right.
\end{equation}
where $\omega=\lambda^{\frac{1}{6}}t_{\mathrm{reh}}^{\frac{1}{3}}\alpha$ and $A_h=A\lambda^{-\frac{1}{3}}t_{\mathrm{reh}}^{-\frac{2}{3}}$. Here $h$ is the Higgs field value at the end of inflation, and we have omitted the $\mathrm{sgn}(h)$ since $\Gamma$ is an even function of $h_\mathrm{reh}$. Note that to ensure accuracy, for the cases with $t_{\mathrm{reh}}\sim t_\mathrm{c}$, the numerical solution of Eq.~\eqref{eq:KG_higgs} is employed in the following computations. To evaluate the effectiveness of different methods across various parameter regions, we treated $\omega$ as a free parameter, which is related to $\Gamma_0$ through the relations

\begin{equation}
H(t_{\mathrm{reh}})={1}/{(2t_{\mathrm{reh}})}\approx\Gamma_0,~ \omega=\lambda^{\frac{1}{6}}t_{\mathrm{reh}}^{\frac{1}{3}}\alpha~.  
\end{equation}
Since the oscillatory behavior of the Higgs field occurs after $t > t_c \approx {\sqrt{2}}/{(3\sqrt{\lambda}\bar{h})}$,  we mainly focus on the case $\omega>\omega_c=\lambda^{\frac{1}{6}}t_{c}^{\frac{1}{3}}\alpha\sim H_{\mathrm{inf}}^{-1/3}$. For $\omega<\omega_c$, the standard $\delta N$ expansion can yield reliable results.

Note that the Higgs field may attain an average value around $\mathcal{O}(H_{\mathrm{inf}})$ in our observed universe during inflation. To illustrate our calculation,  we consider two typical average values for the Higgs value across our observable universe: $\bar{h}=~2H_{\mathrm{inf}},~4H_{\mathrm{inf}}$. The numerical results are provided below. 

\subsection{The power spectrum}

\begin{figure}[htbp]
    \centering
    \begin{subfigure}[b]{0.8\textwidth}
        \includegraphics[width=\textwidth]{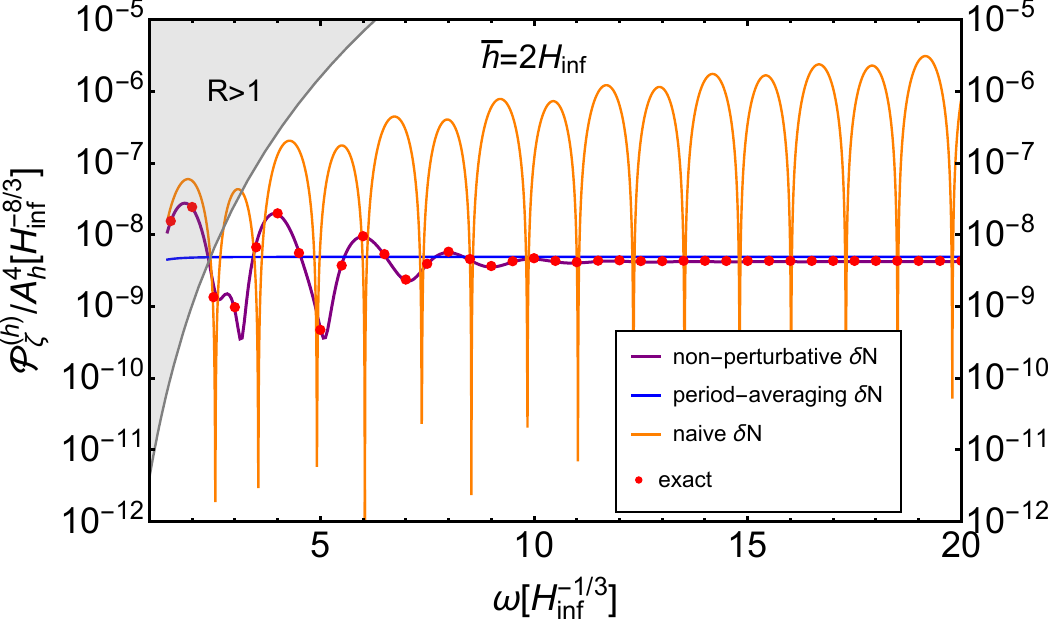} 
    \end{subfigure}
    \vspace*{.2cm}
    
    \begin{subfigure}[b]{0.8\textwidth}
        \includegraphics[width=\textwidth]{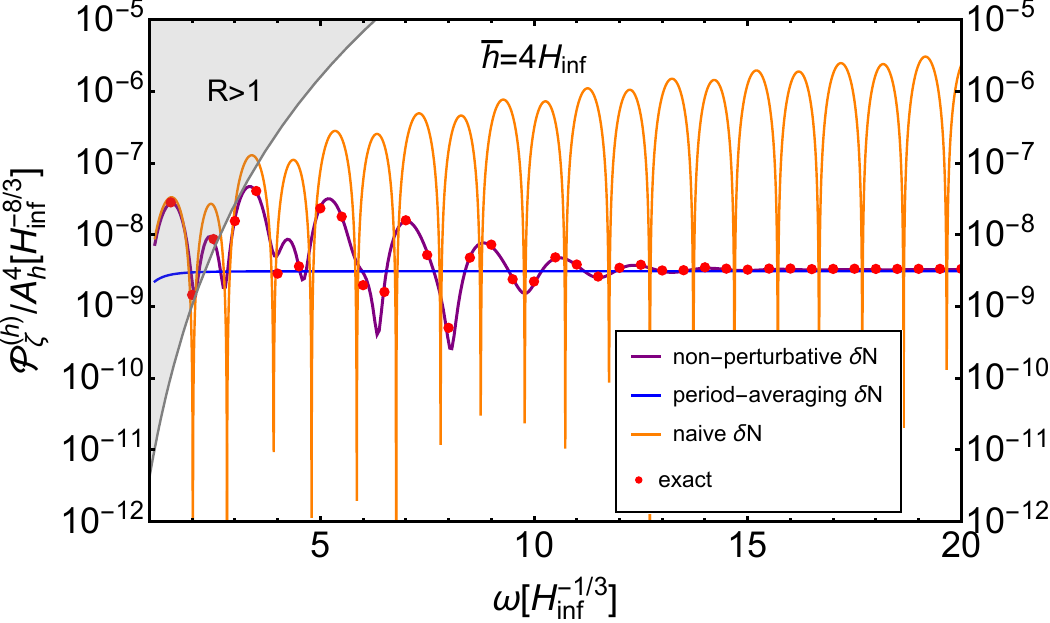} 
    \end{subfigure}    
    \caption{\label{fig:power_spectrum}The power spectrum of curvature perturbation calculated in four different methods in the $(\omega, \mathcal{P}_\zeta^{(h)}/A_h^4)$ plane.}
\end{figure}

\begin{figure}[htbp]
    \centering
    \begin{subfigure}[b]{0.8\textwidth}
        \includegraphics[width=\textwidth]{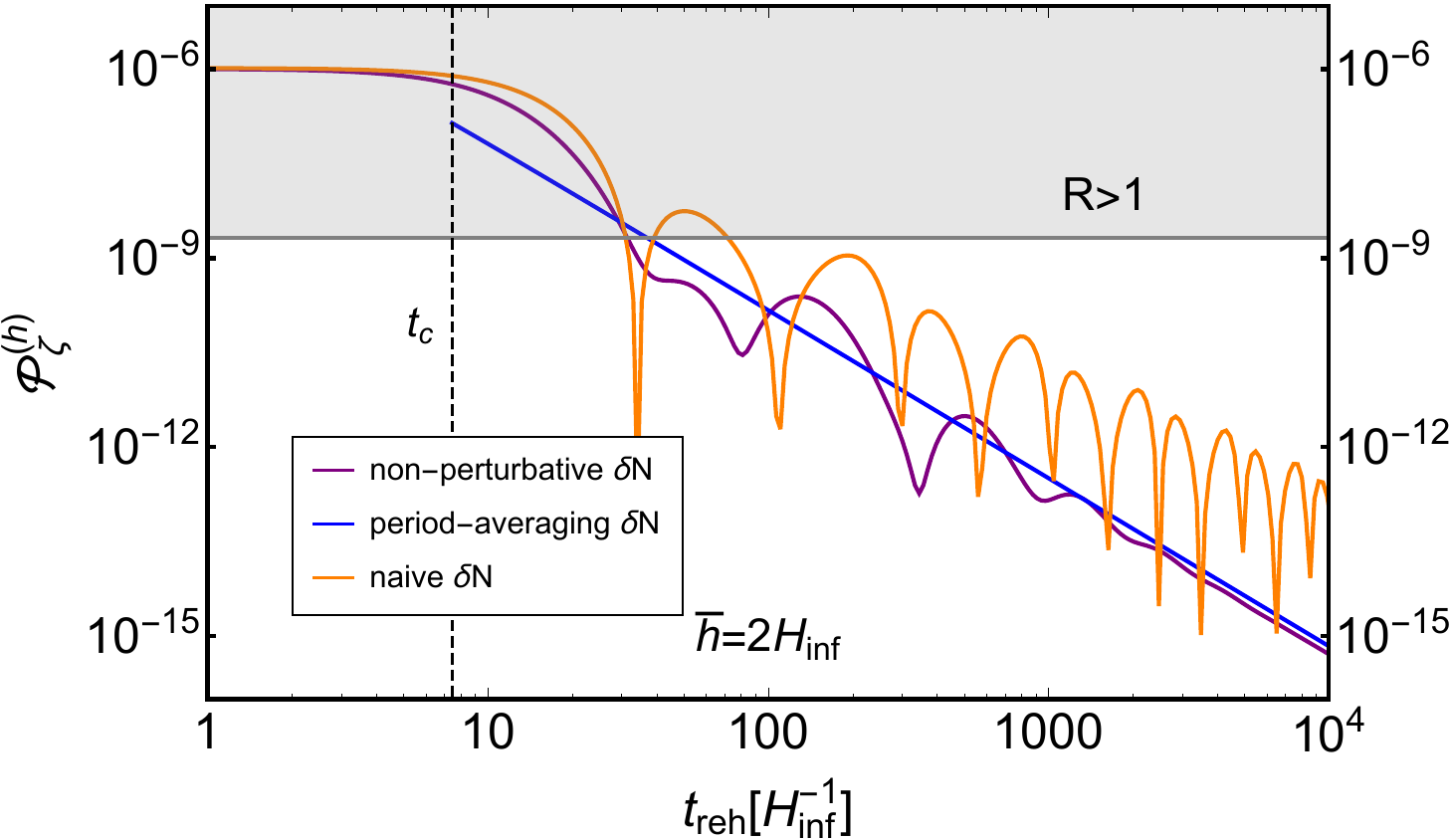} 
    \end{subfigure}
     \vspace*{.5cm}
     
    \begin{subfigure}[b]{0.8\textwidth}
        \includegraphics[width=\textwidth]{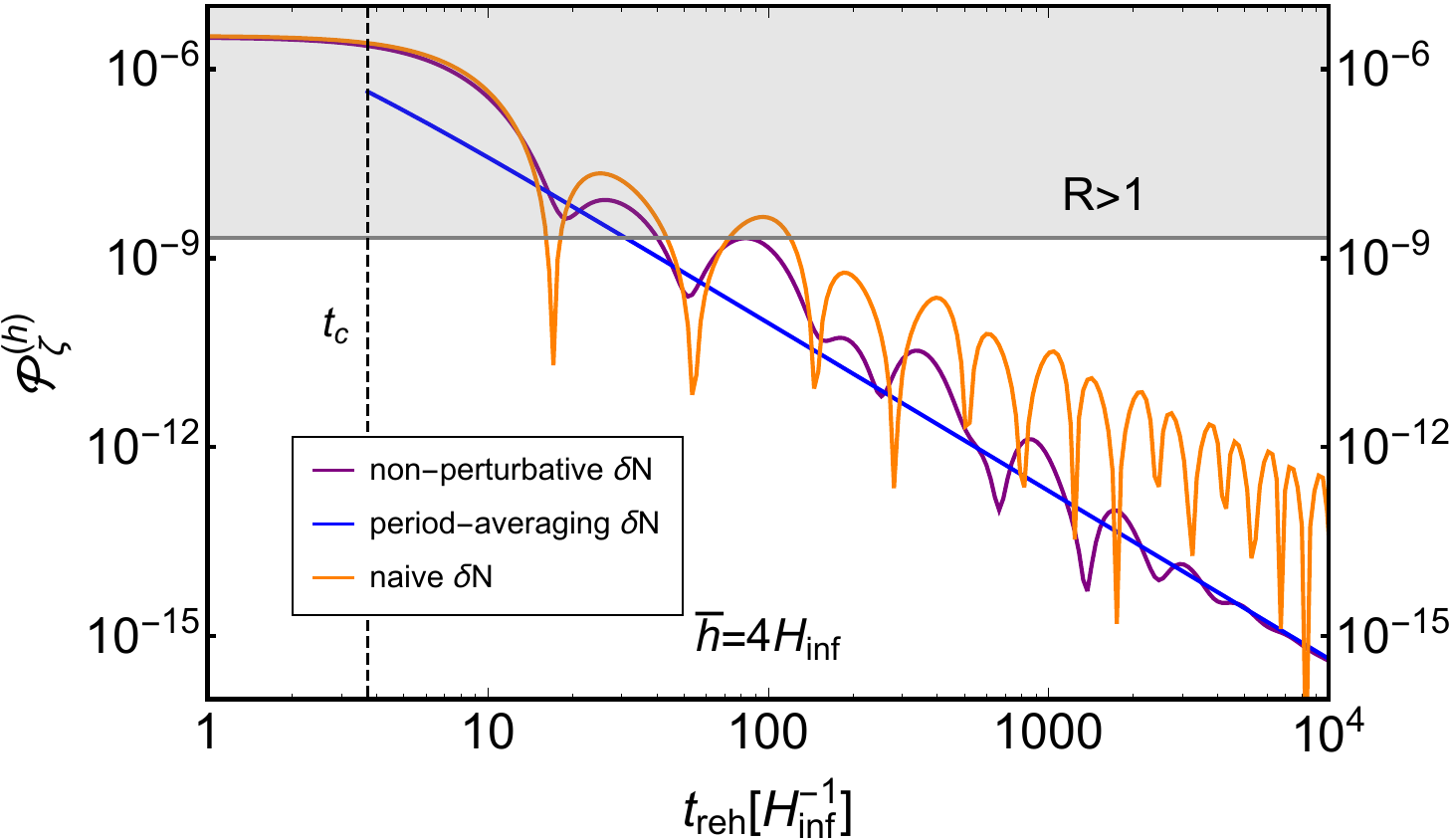} 
    \end{subfigure}    
    \caption{\label{fig:power_spectrum_treh}The power spectrum of curvature perturbation calculated in three different methods in the $(t_\mathrm{reh}, \mathcal{P}_\zeta^{(h)})$ plane.}
\end{figure}

The dimensionless power spectrum of curvature perturbations from Higgs modulated reheating, $\mathcal{P}^h_\zeta$, is calculated using three different methods: the period-averaging method, the exact method, and the non-perturbative $\delta N$ method, which correspond to Eqs. (\ref{eq:power_PA}), (\ref{eq:power_exact}), and (\ref{eq:reduced_spectrum_NP}), respectively. The numerical results are shown in Fig.~\ref{fig:power_spectrum} in the $(\omega, \mathcal{P}_\zeta^{(h)}/A_h^4)$ plane and Fig.~\ref{fig:power_spectrum_treh} in the $(t_\mathrm{reh}, \mathcal{P}_\zeta^{(h)})$ plane, where the upper and lower panels correspond to $\bar h = 2 H_{\mathrm{inf}}$ and $\bar h = 4 H_{\mathrm{inf}}$, respectively. The factor $1/A_h^4$ is introduced in Fig.~\ref{fig:power_spectrum} to ensure a clear and fair comparison, since the power spectrum scales as $A_h^{-4}$, given that $\zeta$ approximately depends on $A_h^{2}$. The range of $\omega$ in Fig.~\ref{fig:power_spectrum} is set to $\omega_\mathrm{c} \leq \omega \leq 20$, corresponding to a maximum reheating time of $t_{\mathrm{reh}} \sim 10^4 H_{\mathrm{inf}}^{-1}$. The results of the period-averaging and non-perturbative $\delta N$ methods are shown as blue and purple lines, respectively. The results of the exact method, shown as red dots, are obtained by Fourier transforming the exact two-point functions using the FFT algorithm. Due to the high computational cost, the power spectrum from the exact method is evaluated only at discrete values of $\omega$, chosen as $\omega = 0.5 n$ with $n = 2, 3, \dots, 40$. Note that all the calculation is estimated in a comoving scale $k_*=100k_\mathrm{min}= 0.045$ Mpc$^{-1}$. The gray regions correspond to $R>1$ and is excluded by the CMB observation. 

As shown in Fig.~\ref{fig:power_spectrum}, the power spectrum $\mathcal{P}_\zeta^{(h)}/A_h^4$ obtained from the period-averaging method remains constant with respect to $\omega$. This is because the $\omega$-dependent terms are already averaged out, eliminating any explicit $\omega$ dependence. In contrast, the power spectrum $\mathcal{P}_\zeta^{(h)}/A_h^4$ from the non-perturbative $\delta N$ method agrees very well with that from the exact method across the entire range of $\omega$. For small $\omega$, both results oscillate around the period-averaged result. As $\omega$ increases, the exact result gradually converges to the period-averaged result. This behavior is expected: at large $\omega$, the statistical average of the field fluctuations spans many oscillation periods, leading to convergence toward the period average. 


From the plot, we find that the period-averaging method provides a good approximation for $\omega \gtrsim 10$ when $\bar h = 2H_{\mathrm{inf}}$ and for $\omega \gtrsim 15$ when $\bar h = 4H_{\mathrm{inf}}$. A naive estimate of the validity range of the period method can be obtained by expanding $h = \bar h + \delta h$ with $\delta h \sim H_{\mathrm{inf}}$, which gives
\begin{equation}
\cos^2(\omega |h|^{1/3}+\theta) \rightarrow \tfrac{1}{2} \left[\cos\left(\tfrac{2}{3}\omega \bar h^{1/3} \delta h + \theta^\prime\right) + 1 \right]~.
\end{equation}
Requiring sizable oscillations,
\begin{equation}
\tfrac{2}{3}\omega \bar h^{-2/3} \delta h \gtrsim 2\pi~,
\end{equation}
leads to $\omega \gtrsim \mathcal{O}(10)$ for $\bar h \sim H_{\mathrm{inf}}$.

For comparison, we also include the naive $\delta N$ expansion, retaining only the leading-order contribution, 
\begin{equation}
\mathcal{P}_{\zeta}^{(h)}= {N^\prime}^2 \frac{H_{\mathrm{inf}}^{2}}{4\pi^{2}}~.
\end{equation}
which is shown as the orange curves in Figs.~\ref{fig:power_spectrum} and \ref{fig:power_spectrum_treh}. We see that $\mathcal{P}_\zeta^{(h)}/A_h^4$ also exhibit oscillations, with amplitudes that grow as $\omega$ increases. The oscillatory behavior arises because $N'$ itself is an oscillatory function of $\omega$, while the growing amplitude reflects the fact that $N'$ contains an overall factor of $\omega$. The naive $\delta N$ expansion works reasonably well for small $\omega$ ($\omega < 2$). But as $\omega$ increases, its deviation from the exact result becomes progressively larger. This indicates that the naive $\delta N$ approach is not reliable for estimating curvature perturbations at large $\omega$. The breakdown occurs because each derivative of $N$ introduces an additional factor of $\omega$, so higher-order terms are not guaranteed to remain small. In principle, one would need to re-sum all orders, which is technically challenging within the naive $\delta N$ framework. By contrast, the non-perturbative $\delta N$ approach naturally performs such a re-summation and remains accurate even at large $\omega$.

In Fig.~\ref{fig:power_spectrum_treh}, we show how the power spectrum varies with different reheating times, $t_{\text{reh}}$, as computed by several methods. The horizontal line labeled $t_c$ indicates the time after which the Higgs field begins to oscillate. This value depends on the Higgs field amplitude, but as a representative timescale, we fix the Higgs field to its mean value $\bar{h}$ when evaluating $t_c$, namely,
$t_c = \frac{\sqrt{2}}{3\sqrt{\lambda}\bar h}$. We find that when $t_{\text{reh}} < t_c$, the power spectrum remains nearly constant, and the results from the naive and non-perturbative $\delta N$ approaches are consistent with each other. This is because, for $t_{\text{reh}} < t_c$, the Higgs field retains its value from the end of inflation, and no significant nonlinear evolution occurs. In contrast, when $t_{\text{reh}} > t_c$, the Higgs field has already decreased and begun oscillating, so the curvature perturbation sourced by Higgs modulated reheating starts to diminish. As $t_{\text{reh}}$ increases, the result from the non-perturbative $\delta N$ method gradually approaches that obtained by the period-averaging method, whereas the naive $\delta N$ method predicts a much larger curvature perturbation, consistent with our earlier discussion.

\subsection{The three-point correlation function }

For the three-point correlation function, the exact result in $x$-space can be directly calculated using Eq.~\eqref{eq:exact_cor_func} by treating $p(h_1,\dots,h_m)$ as a three-dimensional Gaussian distribution with mean $\bar{h}$ and covariance matrix $\Sigma_{ij} = \Sigma(r_{ij}) = \langle \delta h(\mathbf{x}_i)\delta h(\mathbf{x}_j)\rangle$. As mentioned earlier, obtaining the bispectrum in $k$-space is numerically difficult. Since the non-perturbative $\delta N$ method reproduces the exact result for the power spectrum, we expect the three-point correlation function to converge to the exact result as well. More importantly, its functional form, as is provided in Eq.~\eqref{eq:3pt_NP}, is consistent with local-type non-Gaussianity, which can be readily transformed into $k$-space. Note that we have assumed negligible Higgs self-coupling, in which case the non-Gaussianity of curvature perturbation after reheating is generated from the non-linear evolution of the initial Gaussian field perturbation. The Higgs self-coupling could also generate non-Gaussian curvature perturbation, which will be discussed in Sec. \ref{subsec:self_coupling}.

To facilitate comparison between the two methods, we define
\begin{equation}
\mathbf{r}_{ij} = \mathbf{x}_i - \mathbf{x}_j~.
\end{equation}
By translation invariance, the three-point correlation function depends only on $r_{12}, r_{13},$ and $r_{23}$, the side lengths of the corresponding triangle. In our analysis, we take $r_{12}$ to be the largest scale and fix it at $r_{12} = 0.045~\text{Mpc}^{-1}$. We then present the three-point correlation function in terms of the triangle ratios $r_{13}/r_{12}$ and $r_{23}/r_{12}$, which vary within the range (0,1].

\begin{figure}[htbp]
    \centering
    \begin{subfigure}[b]{0.48\textwidth}
        \includegraphics[width=\textwidth]{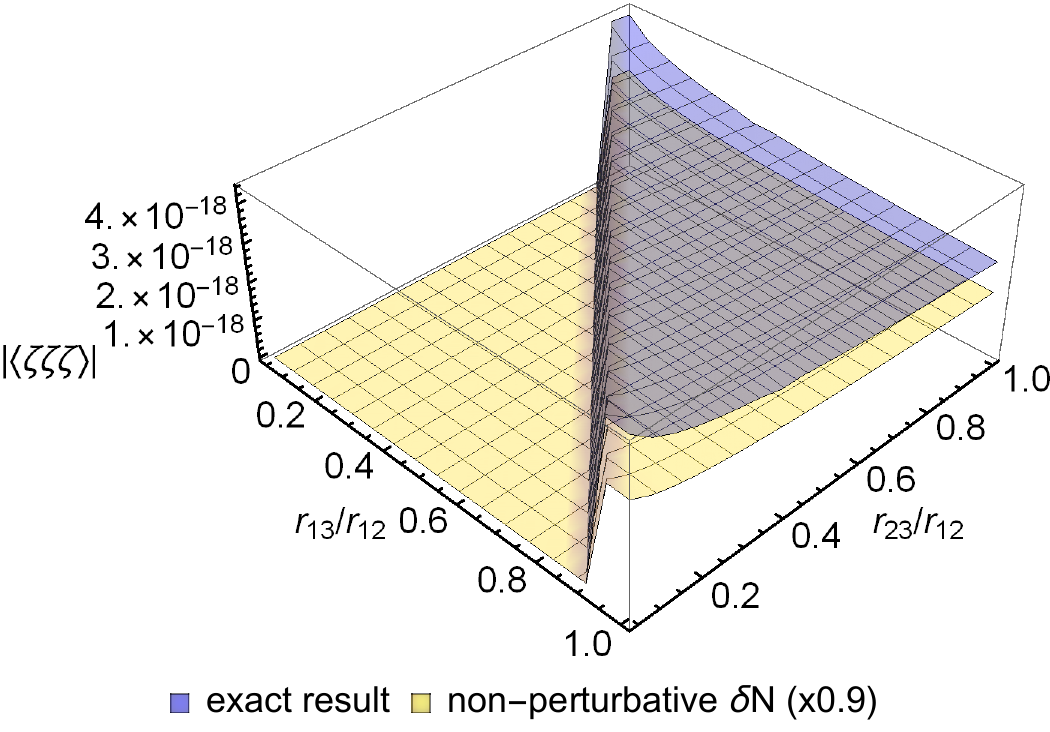} 
        \caption{\label{fig:3pt11}$\bar{h}=2H_{\mathrm{inf}}$, $\omega=5$}
    \end{subfigure}
    \begin{subfigure}[b]{0.48\textwidth}
        \includegraphics[width=\textwidth]{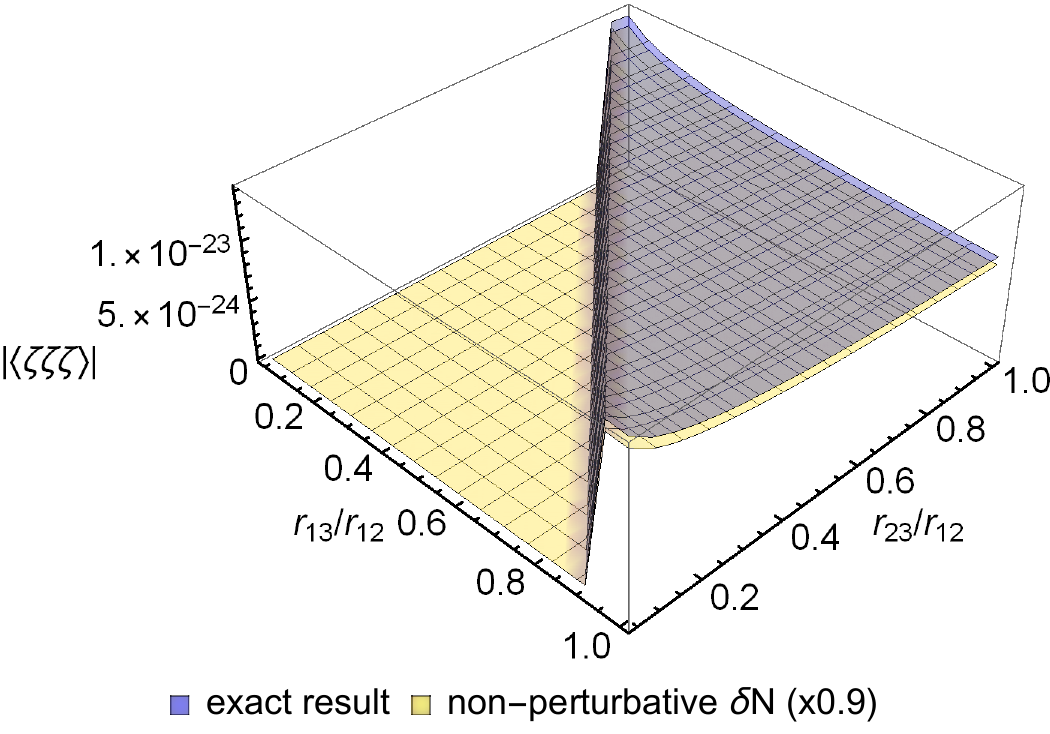} 
        \caption{\label{fig:3pt12}$\bar{h}=2H_{\mathrm{inf}}$, $\omega=15$}
    \end{subfigure}
     \vspace*{.5cm}
     
    \begin{subfigure}[b]{0.48\textwidth}
        \includegraphics[width=\textwidth]{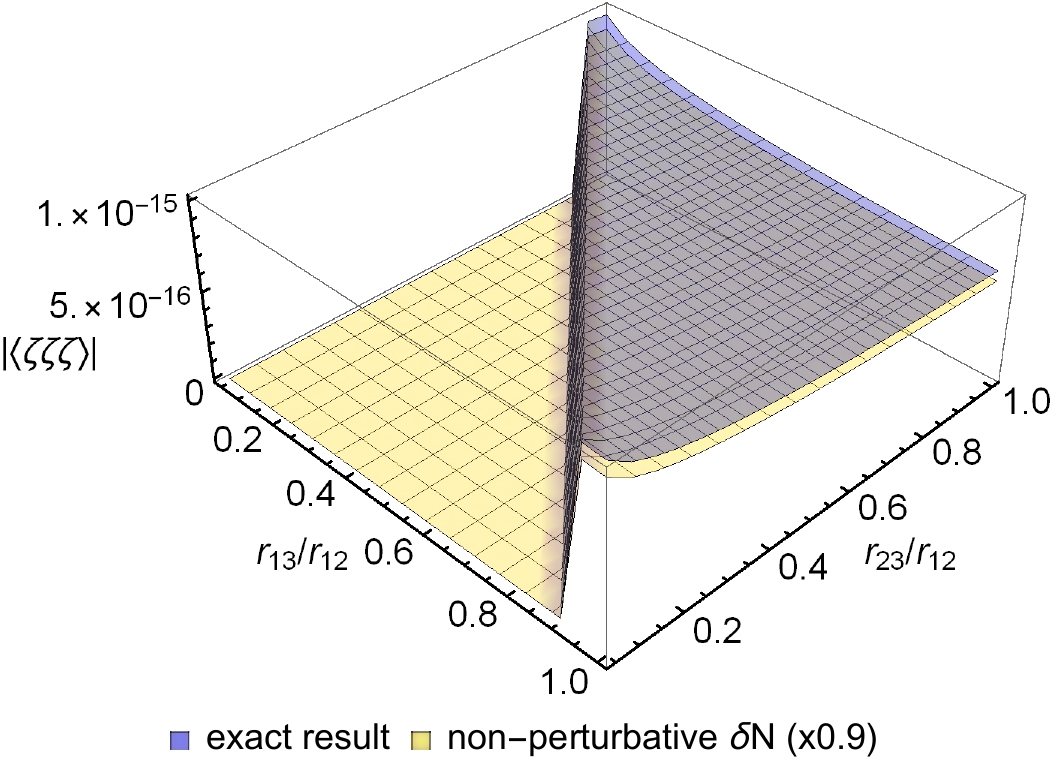} 
        \caption{\label{fig:3pt21}$\bar{h}=4H_{\mathrm{inf}}$, $\omega=5$}
    \end{subfigure}
    \begin{subfigure}[b]{0.48\textwidth}
        \includegraphics[width=\textwidth]{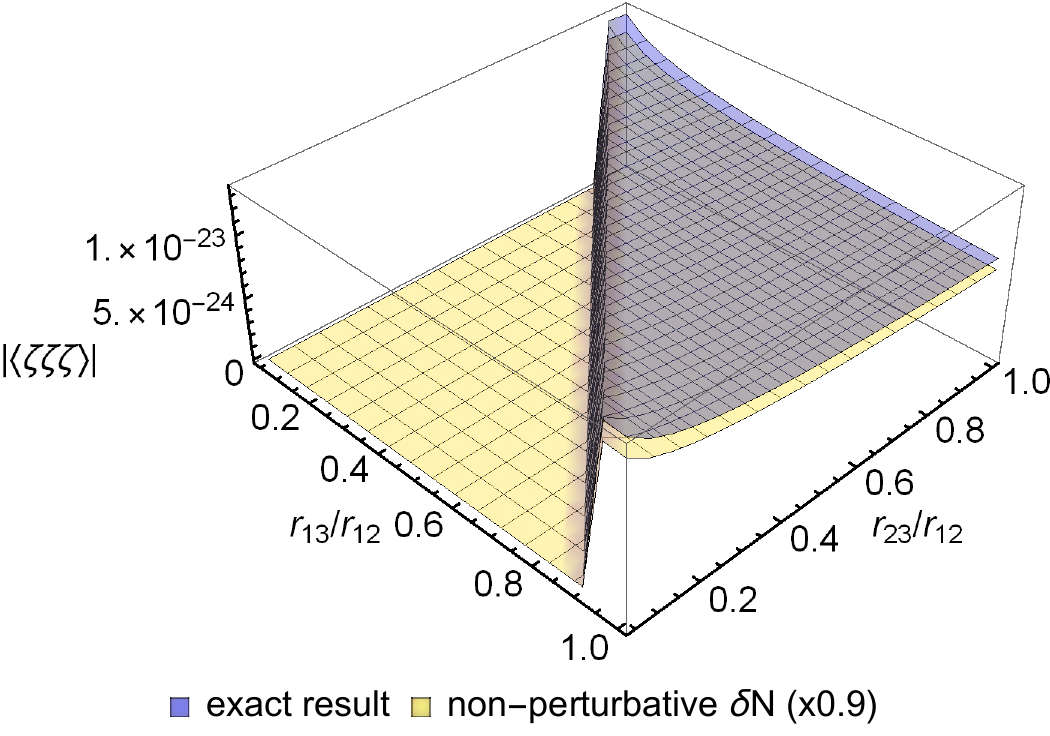} 
        \caption{\label{fig:3pt22}$\bar{h}=4H_{\mathrm{inf}}$, $\omega=15$}
    \end{subfigure}
    \caption{\label{fig:3pt}The three-point function of curvature perturbation calculated using exact and non-perturbative $\delta N$ methods.}
\end{figure}

The three-point correlation functions computed using the exact method and the non-perturbative $\delta N$ method are shown as the blue and yellow surfaces in Fig.~\ref{fig:3pt}. The upper and lower panels correspond to $\bar h = 2 H_{\mathrm{inf}}$ and $\bar h = 4 H_{\mathrm{inf}}$, respectively. For illustration, we present results for two different values of $\omega$: the left panels use $\omega = 5$, and the right panels use $\omega = 15$. To facilitate a clearer visual comparison, the results obtained from the non-perturbative $\delta N$ method have been rescaled by a factor of 0.9.

We find that all the plots exhibit good agreement between the exact and non-perturbative $\delta N$ results.
In Fig.~\ref{fig:3pt11}, although the numerical difference appears relatively large, the deviation remains below $40\%$. We have verified that this discrepancy mainly arises from higher-order corrections in the non-perturbative $\delta N$ method. For the remaining three plots, the difference between the two methods is less than $4\%$. This confirms that the non-perturbative $\delta N$ method provides a reliable estimate of the non-Gaussianity shape. These results demonstrate that the non-perturbative $\delta N$ approach yields accurate predictions not only for the power spectrum but also for higher-order correlation functions. In particular, they indicate that the bispectrum shape is predominantly of the local type.

\begin{figure}[htbp]
    \centering
    \begin{subfigure}[b]{0.8\textwidth}
        \includegraphics[width=\textwidth]{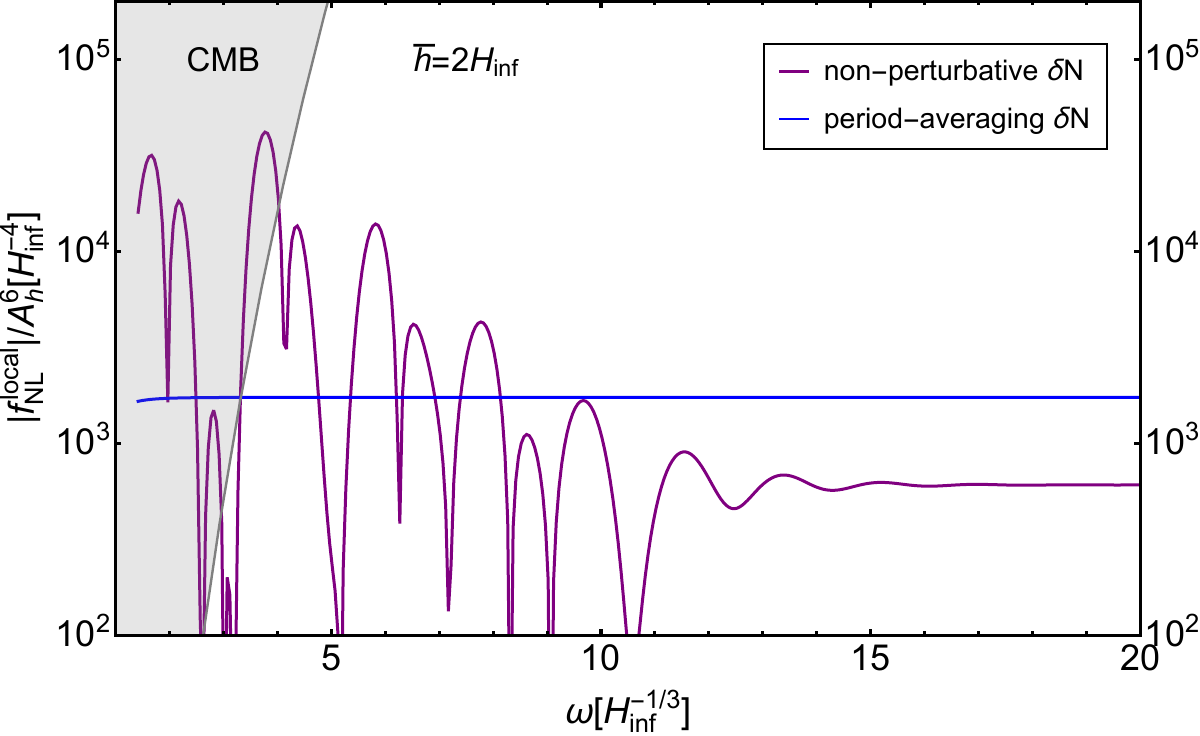} 
    \end{subfigure}
     \vspace*{.5cm}
     
    \begin{subfigure}[b]{0.8\textwidth}
        \includegraphics[width=\textwidth]{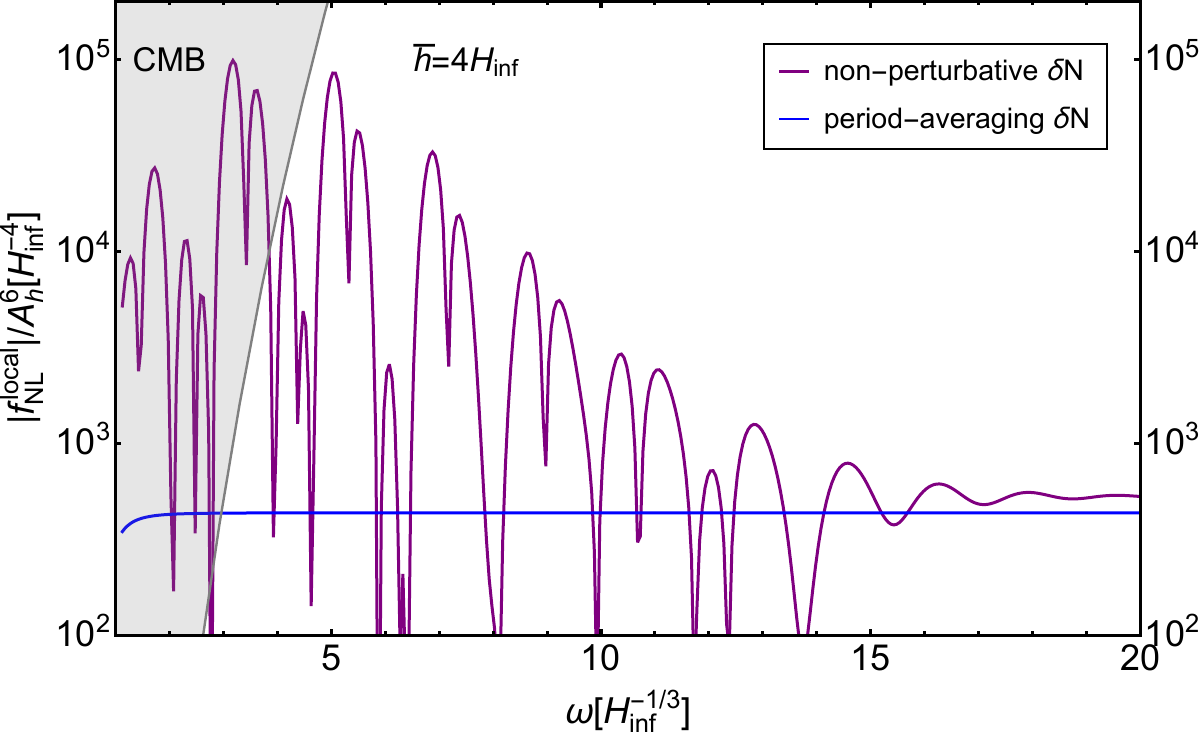} 
    \end{subfigure}    
    \caption{\label{fig:bispectrum}The reduced bispectrum of curvature perturbation calculated in period-averaging and non-perturbative $\delta N$ method.}
\end{figure}

From the above analysis, we find that the bispectrum shape is predominantly of the local type. 
We then could calculate $f^{\mathrm{local}}_{\mathrm{NL}}$ using both the period-averaging and non-perturbative $\delta N$ methods, as given in Eqs.~(\ref{eq:fnl_PA}) and (\ref{eq:reduced_spectrum_NP}). The numerical results are shown in Fig.~\ref{fig:bispectrum}, where the blue and purple lines represent the two methods in the $(\omega, |f^\mathrm{local}_{\mathrm{NL}}|/A_h^6)$ plane. The gray region correspond to $|f_\mathrm{NL}^\mathrm{local}|>11$, which is excluded by the Planck 2018 result at $95\%$ confidence level \cite{Planck:2019kim}. The upper and lower panels correspond to $\bar h = 2 H_{\mathrm{inf}}$ and $\bar h = 4 H_{\mathrm{inf}}$, respectively.  

Since the three-point correlation function scales as $A_h^{-6}$, we include the factor $1/A_h^6$ for comparison and clarity of presentation. 
The value of $|f^\mathrm{local}_{\mathrm{NL}}|/A_h^6$ obtained from the period-averaging method remains constant with respect to $\omega$, as expected, since the $\omega$-dependent terms are averaged out.  In contrast, the non-perturbative $\delta N$ result exhibits oscillatory behavior at small $\omega$ but gradually converges to a constant as $\omega$ increases.  This trend is natural: for large $\omega$, the statistical average of the field spans many oscillation periods, effectively averaging out the oscillatory component associated with $\omega$. 
We also find that in the large-$\omega$ regime, the results from the non-perturbative $\delta N$ and period-averaging methods agree within an overall factor of $\mathcal{O}(1)$.

\subsection{The $\lambda$ dependence}

\begin{figure}[htbp]
    \centering
    \begin{subfigure}[b]{0.8\textwidth}
        \includegraphics[width=\textwidth]{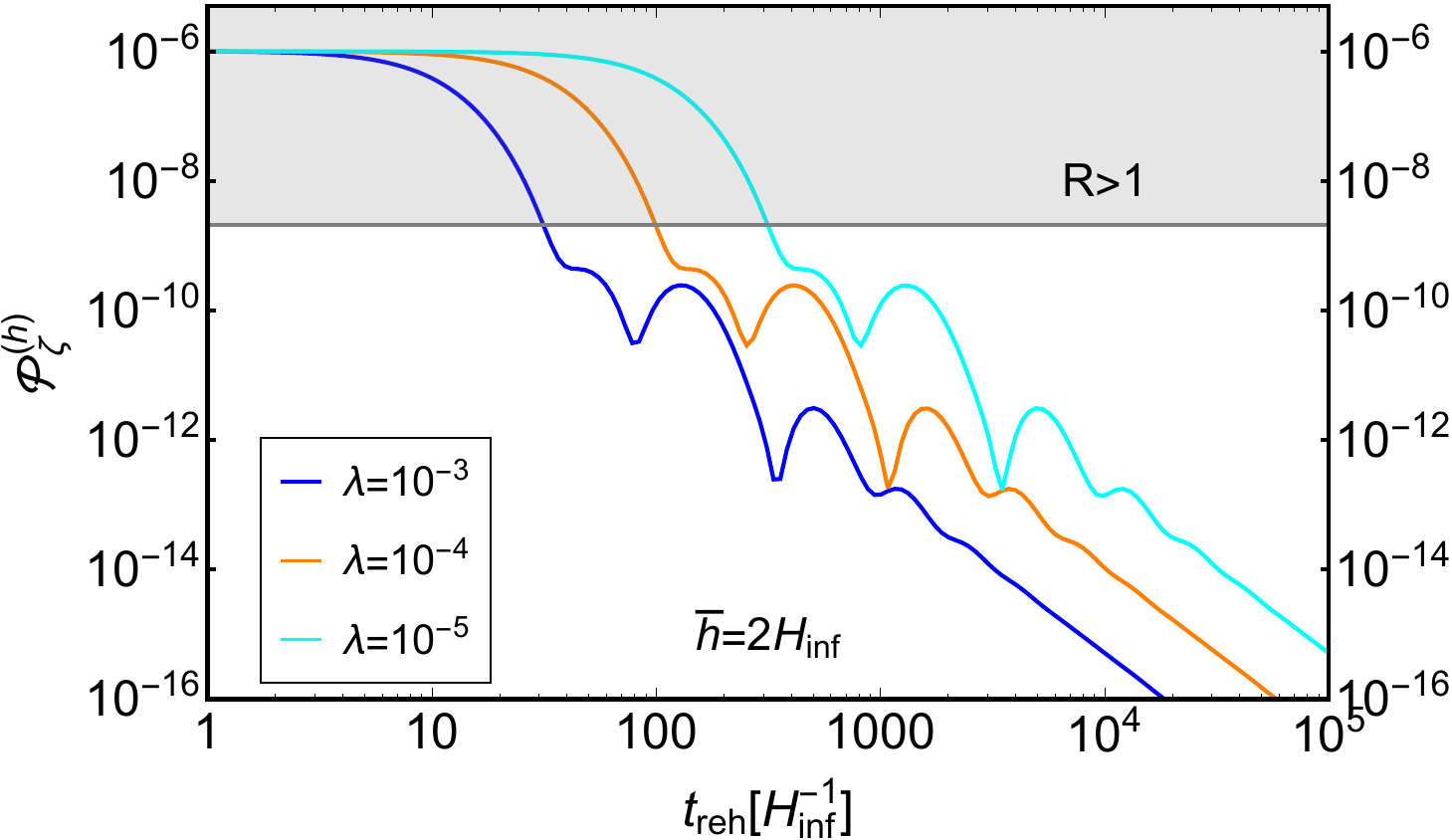} 
    \end{subfigure}
     \vspace*{.5cm}
     
    \begin{subfigure}[b]{0.8\textwidth}
        \includegraphics[width=\textwidth]{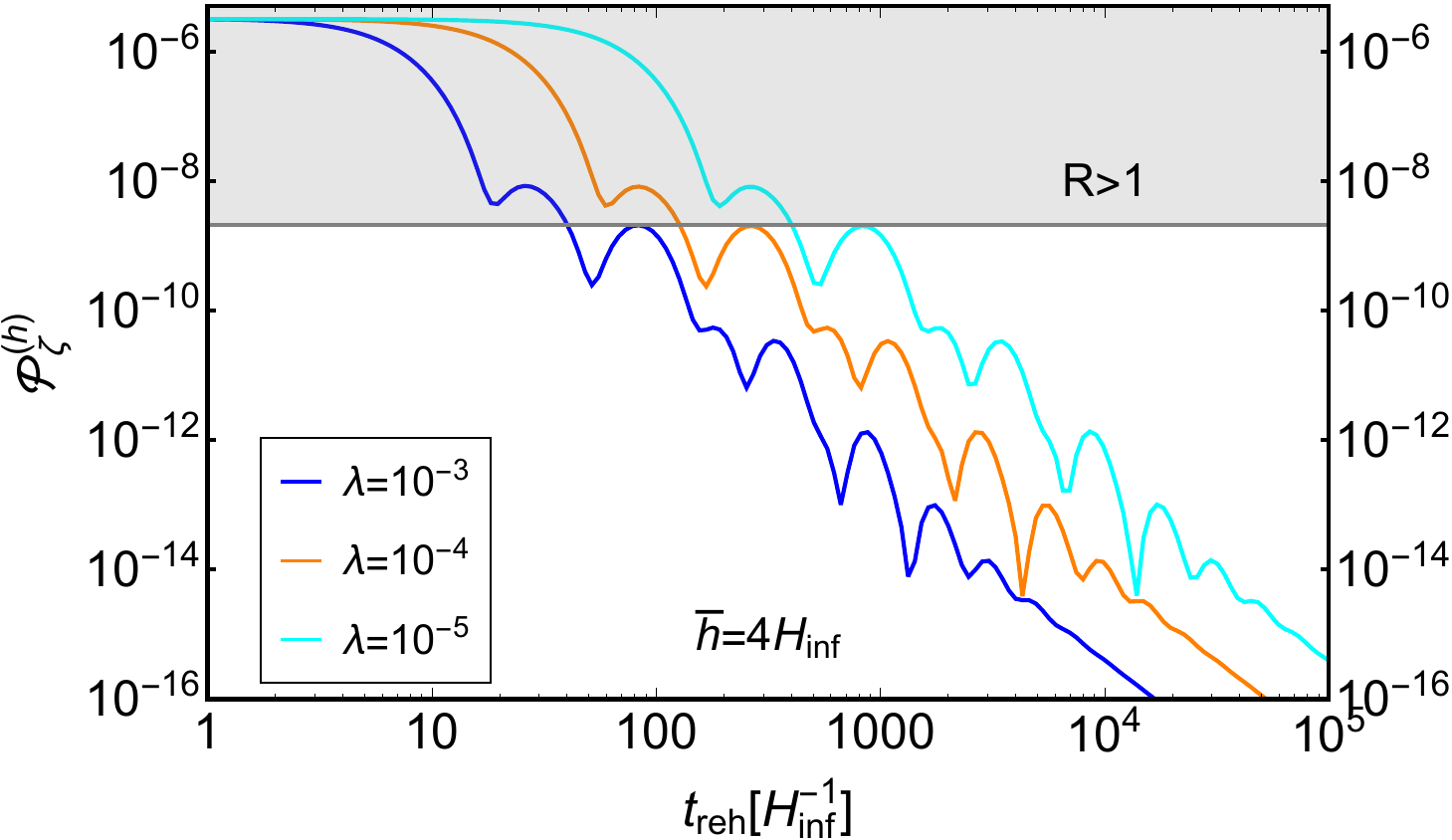} 
    \end{subfigure}    
    \caption{\label{fig:power_lambda}The power spectrum of curvature perturbation with different $\lambda$ calculated in non-perturbative $\delta N$ method.}
\end{figure}

In this subsection, we discuss the dependence of the power spectrum and bispectrum on the Higgs self-coupling $\lambda$. We calculate the power spectrum using the non-perturbative $\delta N$ method, choosing $\lambda = 10^{-3},~10^{-4},~\text{and}~10^{-5}$, as shown in Fig.~\ref{fig:power_lambda}, which correspond to the blue, orange, and cyan lines, respectively. We find that for smaller $\lambda$, while keeping the same $\bar{h}$ value, the resulting curvature perturbation becomes larger. The reason is straightforward: the decay factor of the Higgs amplitude, $A_h$, depends on $\lambda$ as
\begin{equation}
A_h = A\lambda^{-1/3} t_{\mathrm{reh}}^{-2/3}~.
\end{equation}
For smaller $\lambda$, the Higgs field decays more slowly, thereby enhancing the effect of Higgs modulated reheating. In addition, for a smaller $\lambda$, the constant region for the curvature perturbation becomes longer, which is indicated by the relation 
$t_{\mathrm{c}}= \frac{\sqrt{2}}{3\sqrt{\lambda}h_{\mathrm{inf}}}$.


\begin{figure}[htbp]
    \centering
    \begin{subfigure}[b]{0.8\textwidth}
        \includegraphics[width=\textwidth]{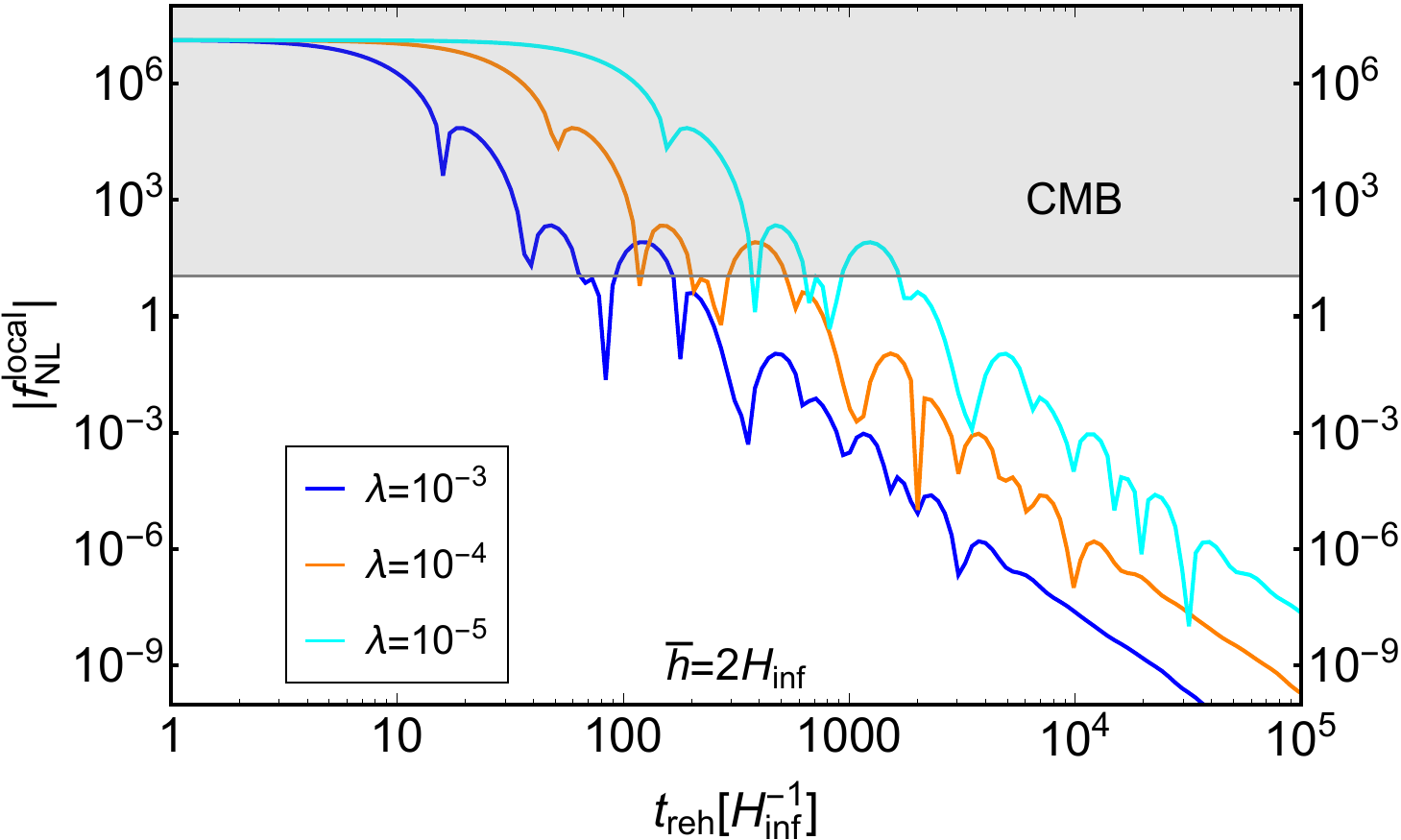} 
    \end{subfigure}
     \vspace*{.5cm}
     
    \begin{subfigure}[b]{0.8\textwidth}
        \includegraphics[width=\textwidth]{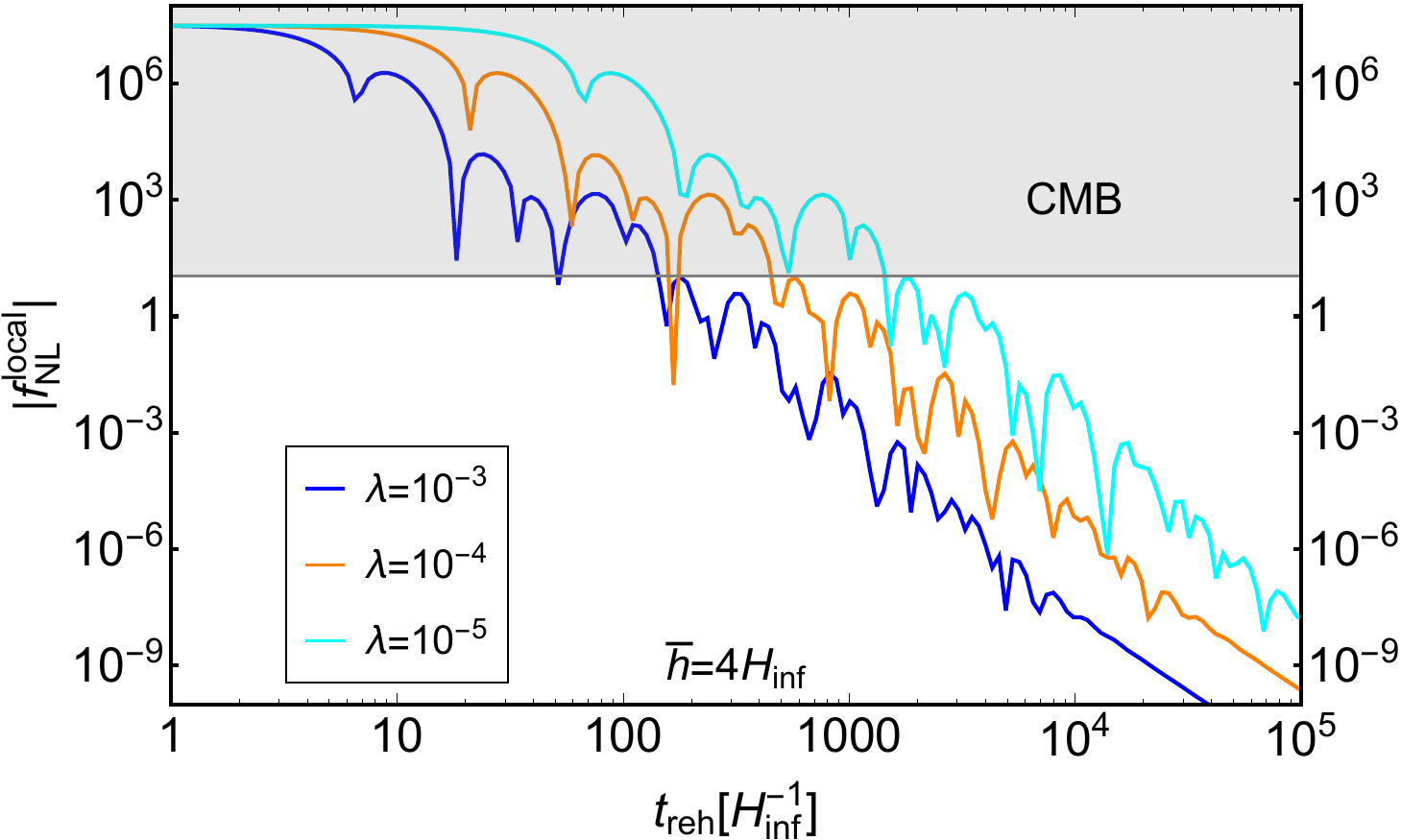} 
    \end{subfigure}    
    \caption{\label{fig:fnl_nonlinear_lambda}The reduced bispectra of curvature perturbation from nonlinear effect with different $\lambda$ calculated in non-perturbative $\delta N$ method.}
\end{figure}

As shown in Fig.~\ref{fig:fnl_nonlinear_lambda}, we compute the bispectrum arising from the non-linear effects using the non-perturbative $\delta N$ method, with $\lambda$ chosen as $10^{-3},~10^{-4},~\text{and}~10^{-5}$, corresponding to the blue, orange, and cyan lines, respectively. Similarly, a smaller Higgs self-coupling leads to a larger bispectrum generated by Higgs modulated reheating.

\subsection{The bispectrum from Higgs self-coupling}
\label{subsec:self_coupling}

\begin{figure}[htbp]
    \centering
    \begin{subfigure}[b]{0.8\textwidth}
        \includegraphics[width=\textwidth]{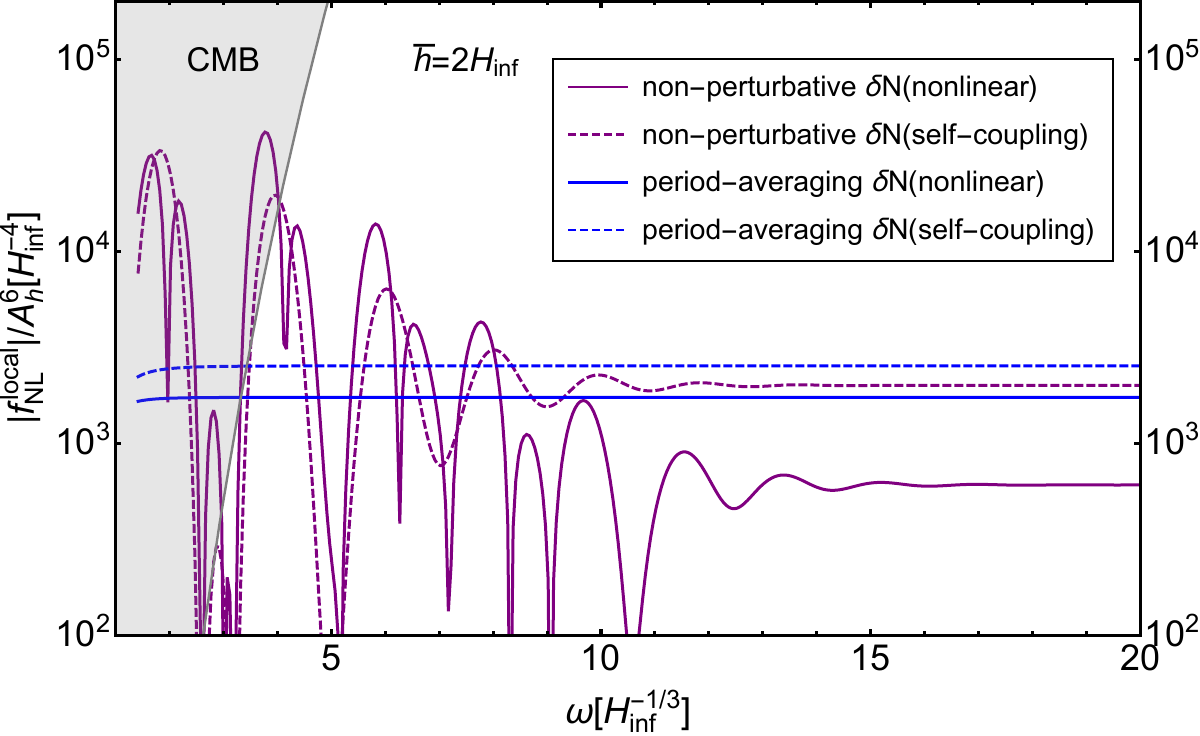} 
    \end{subfigure}
     \vspace*{.5cm}
     
    \begin{subfigure}[b]{0.8\textwidth}
        \includegraphics[width=\textwidth]{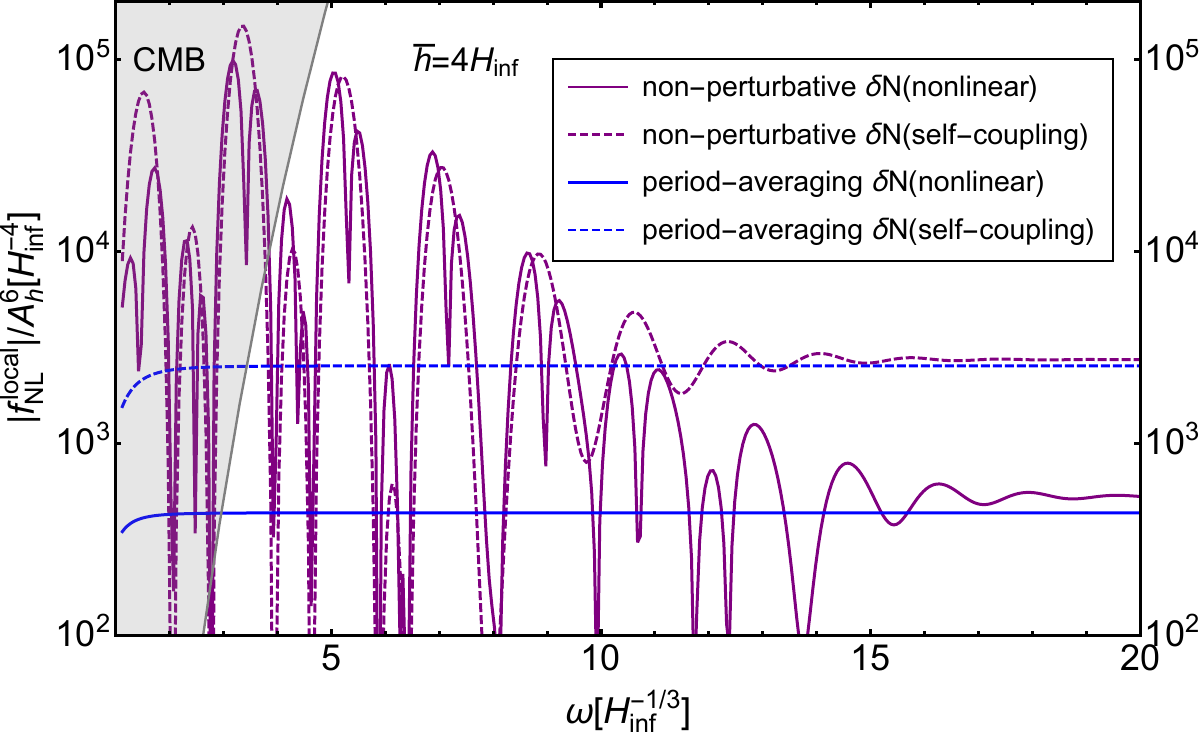} 
    \end{subfigure}    
    \caption{\label{fig:bispectrum_self_coupling}The local bispectra of curvature perturbation from nonlinear effect and Higgs self-coupling with $\lambda=10^{-3}$ calculated in period-averaging and non-perturbative $\delta N$ method.}
\end{figure}

In the previous calculation, we assumed that the joint probability distribution is approximately Gaussian. However, the Higgs self-coupling directly induces non-Gaussianity in the probability distribution. To account for this effect, we include the self-coupling–induced non-Gaussianity,
\begin{equation}
\begin{aligned}
\langle\delta h(\mathbf{x}_{1})\delta h(\mathbf{x_{2}})\delta h(\mathbf{x_{3}})\rangle&=\alpha(r_{12},r_{23},r_{31})~,
\end{aligned}
\end{equation}
The Fourier transform of this expression yields the bispectrum induced by the Higgs self-coupling in momentum space,
\begin{equation}
\begin{aligned}
\langle\delta h_{\mathbf{k}_1}\delta h_{\mathbf{k}_2}\delta h_{\mathbf{k}_3}\rangle&=(2\pi)^3\delta(\mathbf{k}_1+\mathbf{k}_2+\mathbf{k}_3)B_{\delta h}(k_1,k_2,k_3)~,\\
\end{aligned}    
\end{equation}
where $B_{\delta h}(k_1,k_2,k_3)$ can be calculated using the Schwinger–Keldysh path integral formalism~\cite{Weinberg:2005vy,Weinberg:2006ac,Chen:2017ryl,You:2024hit},
\begin{equation}
\label{3pth_SM1}
\begin{aligned}
B_{\delta h}(k_1,k_2,k_3)
&= \frac{1}{\,2\,}\lambda\hp \bar{h} H^2\biggl\{\!\hsm 
\left(\!\frac{1}{\,k_1^3k_2^3\,}\!+\!\mathrm{2\,perm.}\!\right)\!\!\left(\!\!-N_e\!+\!\gamma\!-\!\frac{4}{\,3\,}\!\right)
\\
& \hspace*{19mm}
+\!\frac{1}{\,k_1^2k_2^2k_3^2\,} \!-\!\left(\!\!\frac{1}{\,k_1 k_2^2 k_3^3\,}
\!+\hsm\mathrm{5\,perm.}\!\right )\!\hsm\biggr\} \hp,
\end{aligned}
\end{equation}
with $\gamma\simeq0.577$ the Euler–Mascheroni constant and $N_e\simeq60$. Since the field perturbations generated during inflation are nearly Gaussian, it is appropriate to expand the distribution function $p(h(\mathbf{x}_{1}),\cdots,h(\mathbf{x}_{m}))$ using a Gauss–Hermite series~\cite{Mulryne:2009kh,Mulryne:2010rp,Imrith:2018uyk}: \begin{equation} p(\{h(\mathbf{x}_{i})\})=p_{\mathrm{G}}(\{h(\mathbf{x}_{i})\})\left(1+\frac{A_{im}^{-1}A_{jn}^{-1}A_{kl}^{-1}\alpha_{mnl}H_{ijk}(\{z_i\})}{6}+\cdots\right),
\end{equation} 
where $p_{\mathrm{G}}$ is the multivariate Gaussian distribution with covariance matrix $\Sigma_{ij}\equiv\langle\delta h(\mathbf{x}_{i})\delta h(\mathbf{x}_{j})\rangle$. This covariance can be decomposed as $\Sigma_{ij}=A_{ik}A_{jk}$. We also define $\alpha_{ijk}\equiv\langle\delta h(\mathbf{x}_{i})\delta h(\mathbf{x}_{j})\delta h(\mathbf{x}_{k})\rangle$ and $z_i=A^{-1}_{ij}\delta h(\mathbf{x}_{j})$. The basis function $H_{ijk}$ is given by a generalized Rodrigues’ formula,
\begin{equation} 
H_{ijk}=(-1)^3\exp\!\left(\frac{z_l z_l}{2}\right)\frac{\partial^3}{\partial z_{i}\partial z_{j}\partial z_{k}}\exp\!\left(-\frac{z_m z_m}{2}\right)~.
\end{equation} 
The three-point function of the curvature perturbation can then be estimated as 
\begin{equation} 
\label{eq:3pt_NP-self}
\langle\zeta_h(\mathbf{x}_1)\zeta_h(\mathbf{x}_2)\zeta_h(\mathbf{x}_3)\rangle = \tilde{N}^{\prime3} \alpha(r_{12},r_{23},r_{31}) + \tilde{N}^{\prime\prime}\tilde{N}^{\prime2}\left[\Sigma(r_{12})\Sigma(r_{23})+\mathrm{cyclic}\right]~. 
\end{equation}
In momentum space, the bispectrum obtained in the non-perturbative $\delta N$ method is 
\begin{equation} B^{(h)}_\zeta(k_1,k_2,k_3)=\tilde{N}^{\prime3} B_{\delta h}(k_1,k_2,k_3)+\tilde{N}^{\prime2}\tilde{N}^{\prime\prime}\left[P_{\delta h}(k_{1})P_{\delta h}(k_{2})+\mathrm{cyclic}\right]~. 
\end{equation} 
Focusing on the local-type bispectrum, the nonlinearity parameter $f_\mathrm{NL}^\mathrm{local}$ satisfies 
\begin{equation} f_\mathrm{NL}^\mathrm{local}=\frac{5}{3}\frac{2\lambda\bar{h}}{H_\mathrm{inf}^2\tilde{N}^\prime}\left(-N_e+\gamma-\frac{4}{3}\right)R^4+\frac{5}{6}\frac{\tilde{N}^{\prime\prime}}{\tilde{N}^{\prime2}}R^4~. 
\end{equation}
The first term originates from the Higgs self-coupling, while the second term arises from nonlinear effects. The reduced local bispectrum from the period averaging method can be derived similarly, by replacing the coefficients $\tilde{N}^\prime,~\tilde{N}^{\prime\prime}$ with $z_1,~z_2$.

\begin{figure}[htbp]
    \centering
    \begin{subfigure}[b]{0.8\textwidth}
        \includegraphics[width=\textwidth]{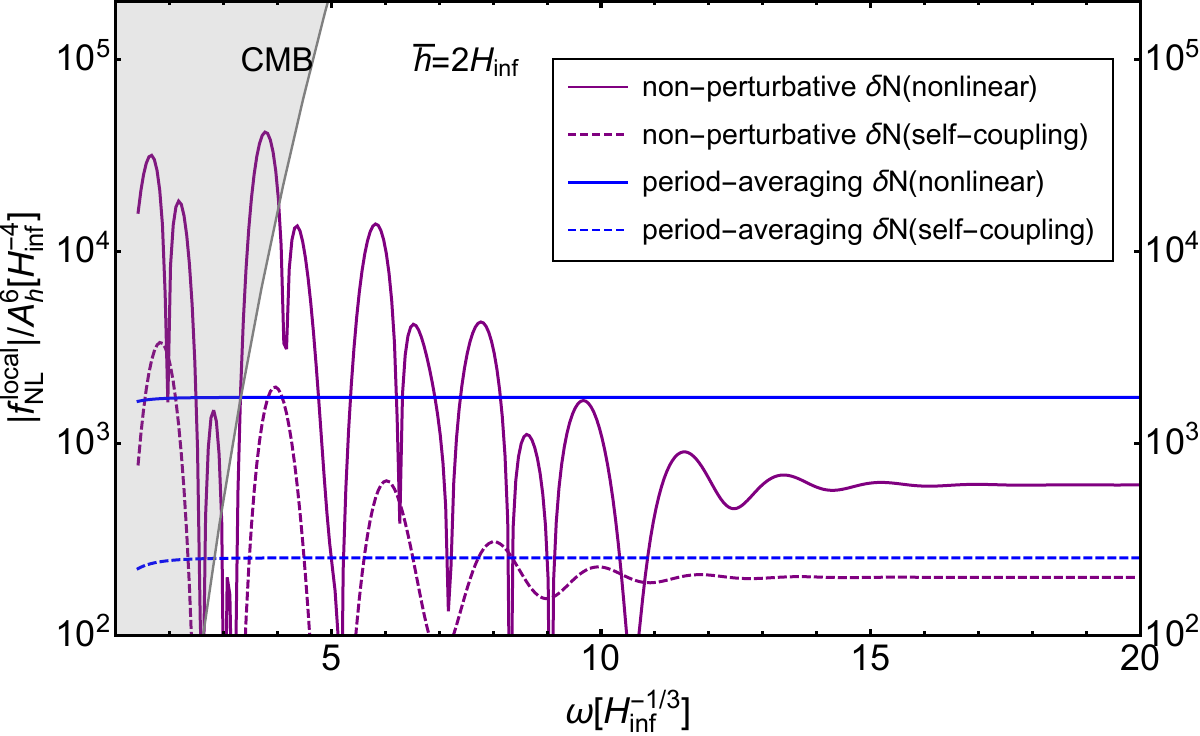} 
    \end{subfigure}
     \vspace*{.5cm}
     
    \begin{subfigure}[b]{0.8\textwidth}
        \includegraphics[width=\textwidth]{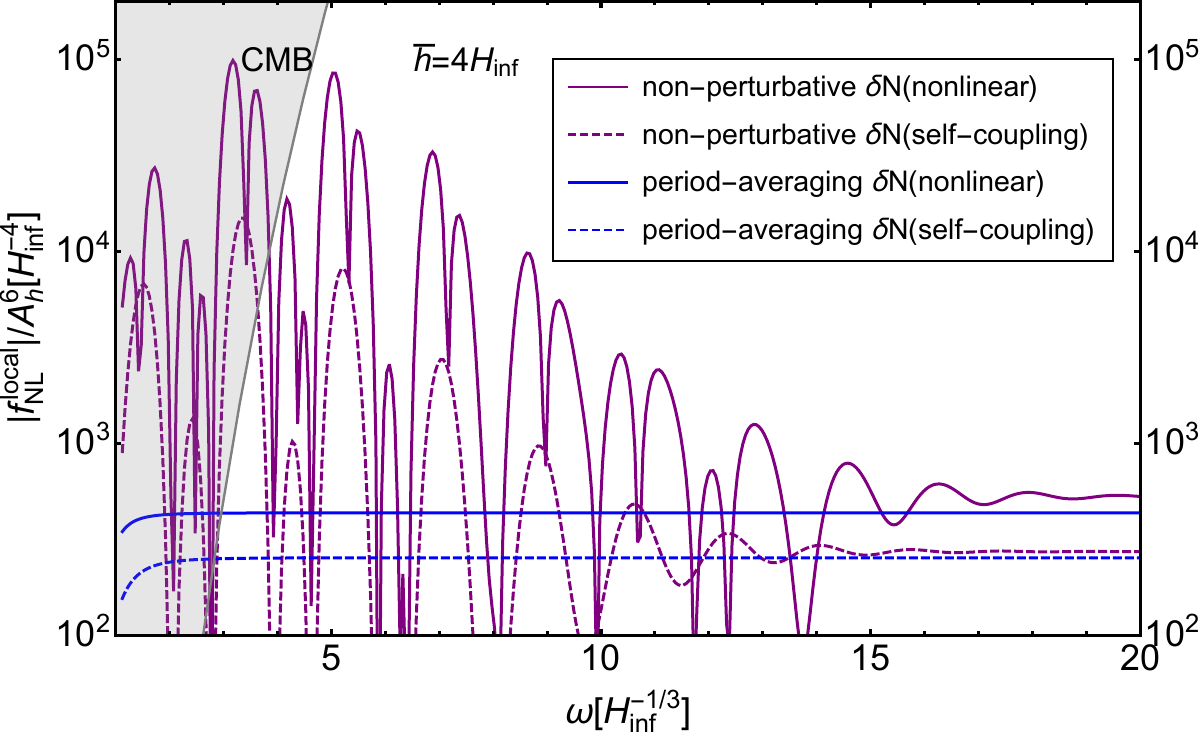} 
    \end{subfigure}    
    \caption{\label{fig:bispectrum_self_coupling2}The local bispectra of curvature perturbation from nonlinear effect and Higgs self-coupling with $\lambda=10^{-4}$ calculated in period-averaging and non-perturbative $\delta N$ method.}
\end{figure}

The numerical results for the local-type bispectrum arising from the Higgs self-coupling with $\lambda=0.001$ and from the nonlinear term are shown in Fig.~\ref{fig:bispectrum_self_coupling}. The upper and lower panels correspond to $\bar{h}=2H_\mathrm{inf}$ and $\bar{h}=4H_\mathrm{inf}$, respectively. The contribution from the Higgs self-coupling is shown by the purple dashed line (non-perturbative $\delta N$ method) and the blue dashed line (period-averaging method). The contribution from nonlinear evolution is shown by the purple solid line (non-perturbative $\delta N$ method) and the blue solid line (period-averaging method). We find that for $\lambda=0.001$, the contributions from the Higgs self-coupling and the nonlinear effect are comparable.

We also present the results for the case $\lambda=10^{-4}$ in Fig.~\ref{fig:bispectrum_self_coupling2}. In this case, the nonlinear contribution dominates over that from the Higgs self-coupling. This behavior is expected since the curvature bispectrum generated by the Higgs self-coupling scales is directly related with $\lambda$. Therefore, for $\lambda < 10^{-3}$, the nonlinear term is expected to dominate.

\subsection{Loop corrections for non-perturbative $\delta N$}
\label{subsubsec:loop_correct}

The non-perturbative $\delta N$ formalism employs an expansion in terms of the spatial cross-correlation function, $\xi(r_{ij})=\frac{\Sigma(r_{ij})}{\Sigma(0)}$. This can be seen from the expansion of the two-point function up to next-to-leading order (NLO):
\begin{equation}
\langle\zeta_h(\mathbf{x}_1)\zeta_h(\mathbf{x}_2)\rangle\simeq \tilde{N}^{\prime2}\Sigma(r_{12})+\frac{1}{2}\tilde{N}^{\prime\prime2}\Sigma^2(r_{12})~,
\end{equation}
where the coefficients $\tilde{N}^\prime$ and $\tilde{N}^{\prime\prime}$ implicitly depend on $\Sigma^{-1}(0)$ and $\Sigma^{-2}(0)$, respectively.
When the condition
\begin{equation}
\left(\frac{\tilde{N}^{\prime\prime}}{\tilde{N}^\prime}\right)^2\Sigma(r_{ij})\ll1~,
\end{equation}
is satisfied, truncating the expansion at leading order (LO) is sufficient for numerical calculations. For the scale of the observable universe, one has $\xi(L_{\mathrm{obs}})\simeq 0.00038$, indicating that the condition is generally satisfied for typical parameter choices. However, this is not the case when $\tilde{N}^\prime \simeq 0$.
Indeed, notice
\begin{equation}
 \tilde{N}^\prime=\Sigma^{-1}(0)\int\mathrm{d}h ~p_\mathrm{G}(h)N\delta h~,
\end{equation}
and since $N$ is an even function of $h$, we obtain $\tilde{N}^\prime=0$ when $\bar h=0$. On the other hand,
\begin{equation}
\tilde{N}^{\prime\prime}=\Sigma^{-2}(0)\int\mathrm{d}h ~p_{\mathrm{G}}(h)(N-\bar{N})\delta h^2~,
\end{equation}
is non-vanishing in this case. Therefore, the NLO correction can become important when $\bar h \approx 0$.
This does not imply a breakdown of the perturbative expansion; rather, it simply reflects that $\tilde{N}^\prime$ happens to vanish in special cases. In what follows, we estimate the LO and NLO contributions to both the power spectrum and the bispectrum.

\begin{figure}[htbp]
	\centering
	\includegraphics[width=0.9\textwidth]{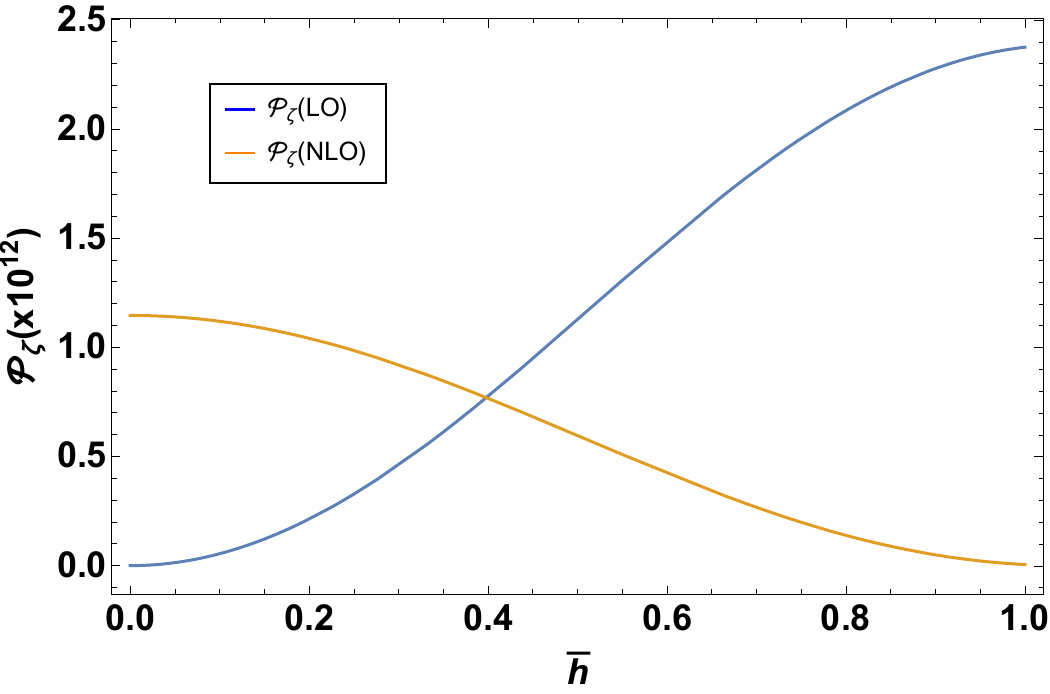}
	\caption{\label{fig:loop}The power spectrum of $\zeta_{h}$ from LO contribution vs. NLO contribution.}
\end{figure}

The leading-order (LO) contribution to the power spectrum is given by
\begin{equation}
\mathcal{P}_{\zeta({\text{LO}})}^{(h)}=\tilde{N}^{\prime2}\mathcal{P}_{\delta h}=\frac{\tilde{N}^{\prime2}H_{\mathrm{inf}}^{2}}{4\pi^{2}}~.
\end{equation}
which is nearly scale-invariant. The next-to-leading-order (NLO) contribution takes the form
\begin{equation}
\mathcal{P}_{\zeta({\text{NLO}})}^{(h)}(k)=\frac{k^3}{2\pi^2}\int d^3r\frac{1}{2}\tilde{N}^{\prime\prime2}\Sigma^2(r)e^{-i\mathbf{k\cdot r}}~.
\end{equation}
Substituting $\Sigma(r)=\int\frac{d^3k}{(2\pi)^3}P_{\delta h}(k) e^{i\mathbf{k\cdot r}}$ into the above expression, we obtain
\begin{equation}
\begin{aligned}
\mathcal{P}_{\zeta({\text{NLO}})}^{(h)}(k)&=\frac{\tilde{N}^{\prime\prime}H^4_{\mathrm{inf}}k^3}{16\pi^2}\int\frac{d^3q}{(2\pi)^3}\frac{1}{q^3}\frac{1}{|\mathbf{k}-\mathbf{q}|^3}\\
&=\frac{\tilde{N}^{\prime\prime}H^4_{\mathrm{inf}}k^3}{4(2\pi)^4}\int^1_{-1} d(\cos\theta)\int dq \frac{1}{q}\frac{1}{(k^2+q^2-2kq\cos\theta)^{3/2}}\\
&=\frac{\tilde{N}^{\prime\prime}H^4_{\mathrm{inf}}k^3}{4(2\pi)^4}\left[\int^{k-k_{\mathrm{min}}}_{k_{\mathrm{min}}}dq\frac{2}{kq(k^2-q^2)}+\int^{k_{\mathrm{max}}}_{k+k_{\mathrm{min}}}dq\frac{2}{q^2(q^2-k^2)}\right]~,
\end{aligned}
\end{equation}
where we have introduced the infrared cutoff $k_{\mathrm{min}}$ on $|\mathbf{k}-\mathbf{q}|$.
For illustration, Fig.~\ref{fig:loop} shows the LO and NLO contributions to the power spectrum $\mathcal P_\zeta^{(h)}$ for different values of $\bar{h}$, with $\omega$ fixed at $5$. For $\bar{h}\lesssim 0.4 H_{\mathrm{inf}}$, the NLO contribution becomes comparable to or even larger than the LO contribution. In the special case $\bar{h}=0$, the LO term $\mathcal{P}^{(h)}_{\zeta(\text{LO})}$ vanishes exactly, and the power spectrum is entirely dominated by the NLO contribution.

For the three-point function, we can expand the  the non-perturbative $\delta N$ treatment to next-to-leading order, which is given by 
\begin{equation}
\langle\zeta_h(\mathbf{x}_1)\zeta_h(\mathbf{x}_2)\zeta_h(\mathbf{x}_3)\rangle\simeq \tilde{N}^{\prime2}\tilde{N}^{\prime\prime}\left(\Sigma(r_{12})\Sigma(r_{13})+\mathrm{cyclic}\right)+\tilde{N}^{\prime\prime3}\Sigma(r_{12})\Sigma(r_{13})\Sigma(r_{23})~.
\end{equation}
Therefore, in the case where $N'\sim 0$, the NLO contribution to the three-point function can become significant, leading to a bispectrum with a shape distinct from the local type. Since our analysis primarily focuses on the bispectrum for $\bar{h}\gtrsim H_{\mathrm{inf}}$, we do not explicitly compare the LO and NLO contributions here. Nevertheless, it should be kept in mind that the NLO term can dominate when $\bar{h}$ is small.

\section{Conclusion}
\label{sec:conclusion}

We systematically explored the curvature perturbations and non-Gaussianities generated by Higgs modulated reheating. We employed different methods — the period-averaging (PA) method, the exact method, and the non-perturbative $\delta N$ formalism — to calculate the power spectrum and bispectrum of curvature perturbations across different regimes of reheating. Our results indicated that the non-perturbative $\delta N$ method provides a reliable estimate for both the power spectrum and higher-point correlation functions, even in the presence of strong oscillatory behavior from the Higgs field.

We found that smaller Higgs self-couplings, characterized by $\lambda$, lead to larger curvature perturbations and more pronounced non-Gaussianity. In particular, we observed that the bispectrum generated by Higgs modulated reheating predominantly exhibits a local-type non-Gaussianity. Our comparison of the period-averaging and non-perturbative $\delta N$ methods highlighted their respective domains of applicability, with the latter being the most versatile across a wider range of reheating times and Higgs field values.

Additionally, we showed that for small values of $\lambda$, the effects of Higgs modulated reheating become more significant, amplifying the curvature perturbations.  We also found that the non-Gaussianity arising from Higgs self-interactions could be significant when $\lambda > 0.001$. These results deepen our understanding of how non-perturbative effects in early Universe dynamics, specifically through the Higgs field, can influence cosmological observables.

\appendix

\section{Non-perturbative $\delta N$ formalism}
\label{app:non_perturb}

The non-perturbative $\delta N$ formalism is developed for the calculation of curvature perturbation to handle cases where one cannot employ a truncated Taylor expansion of $\delta N$ \cite{Suyama:2013dqa, Imrith:2018uyk}. In this section, we review the non-perturbative $\delta N$ method and provide detailed proofs of the formulas used in this paper. We also provide a generalization of this method to compute higher order non-Gaussianities, and the expression for the 4-point correlation function is derived as illustration. For simplicity, we focus on the case where the curvature perturbation is generated by the perturbation of a single field $\chi$. In Sec. \ref{subapp:exact_cor} we provide the exact expression
of the $x$-space correlation functions of curvature perturbation. Then, in Sec. \ref{subapp:gauss_hermite}, we derive the Gauss-Hermite expansion of the Higgs field probability distribution function. In Sec. \ref{subapp:nonpert_coeff} the non-perturbative $\delta N$ coefficients are derived using the Gauss-Hermite expansion and the expansion about spatial correlation functions. 

\subsection{Exact correlation functions}
\label{subapp:exact_cor}

Consider the curvature perturbation $\zeta$ generated by the perturbation of field $\chi$, under the separate universe approximation, $\zeta$ can be expressed as
\begin{equation}
	\zeta(\mathbf{x})=\delta N(\mathbf{x})=N(\chi(\mathbf{x}))-\bar{N}~,
\end{equation}
where $N$ is the local number of e-folds between an initial flat hypersurface and a final uniform density hypersurface, and $\bar{N}=\langle N\rangle$ is the ensemble average of $N$. This constitutes the essence of the $\delta N$ formalism. On the initial flat hypersurface, the field perturbation is given by $\delta\chi(\mathbf{x})=\chi(\mathbf{x})-\bar{\chi}$, which is usually close to Gaussian statistics at horizon-exit. Therefore the statistics of $\delta\chi$ can be specified by the power spectrum and the leading order correction, the bispectrum:
\begin{equation}
\begin{aligned}
\langle\delta\chi_{\mathbf{k_1}}\delta\chi_{\mathbf{k_2}}\rangle&=(2\pi)^3\delta^3(\mathbf{k_1}+\mathbf{k_2})P_\chi(k),\\
\langle\delta\chi_{\mathbf{k_{1}}}\delta\chi_{\mathbf{k_{2}}}\delta\chi_{\mathbf{k_{3}}}\rangle&=(2\pi)^{3}\delta^{3}(\mathbf{k_{1}}+\mathbf{k_{2}}+\mathbf{k_{3}})B_\chi(k_{1},k_{2},k_{3})~.
\end{aligned}
\end{equation}

If $\delta N$ can be Taylor expanded in terms of $\delta\chi$, then the power spectrum and bispectrum of $\zeta$ can be expressed in terms of the Taylor coefficients, $P_\chi$ and $B_\chi$. This is the typical treatment of standard $\delta N$ formalism. However, when the Taylor expansion is invalid, the calculation of correlation functions of $\zeta$ is not that simple. In this case, it is convenient to stay in $x$-space. 

The $x$-space two- and three-point correlation function of $\delta\chi(\mathbf{x})$ are given by
\begin{equation}
\begin{aligned}
\langle\delta\chi(\mathbf{x}_1)\delta\chi(\mathbf{x}_2)\rangle&=\Sigma(r_{12}),\\
\langle\delta\chi(\mathbf{x}_{1})\delta\chi(\mathbf{x_{2}})\delta\chi(\mathbf{x_{3}})\rangle&=\alpha(r_{12},r_{23},r_{31})~,
\end{aligned}
\end{equation}
where $r_{ij}=|\mathbf{x}_i-\mathbf{x}_j|$.

Now we consider the m-point function of $\zeta$. Its exact definition is given by
\begin{equation}
\label{eq:exact_cor}
\begin{aligned}
\langle\zeta_{1}\cdots\zeta_{m}\rangle&=\langle(N_{1}-\bar{N})\cdots(N_{m}-\bar{N})\rangle\\&=\int\mathrm{d}\chi_{1}\cdots\int\mathrm{d}\chi_{m}(N_{1}-\bar{N})\cdots(N_{m}-\bar{N})p(\chi_1,\cdots,\chi_m)~,
\end{aligned}	
\end{equation}
where the subscript $i$ denotes the spatial position, $N_i=N(\mathbf{x}_i),~\chi_i=\chi(\mathbf{x}_i)$, and $p(\chi_1,\cdots,\chi_m)$ is the joint probability distribution function of the m fields $\chi_i$. 
In general cases the calculation of Eq.~\eqref{eq:exact_cor} is rather cumbersome. Fortunately, even when $N$ can not be Taylor expanded, there are still two approximate method available:
\begin{enumerate}
\item[(1)] \textit{Gauss-Hermite expansion:} Since the field perturbations generated during inflation are very close to Gaussian distribution, it is justified to make a Gauss-Hermite expansion for the distribution function $p(\chi_1,\cdots,\chi_m)$ around the Gaussian distribution $p_{\mathrm{G}}(\chi_1,\cdots,\chi_m)$, with non-Gaussian corrections described by higher order cumulants. Such a treatment is also known as cumulant expansion.

\item[(2)] \textit{Expansion in the cross correlation function:} If the power spectrum of $\delta\chi$ is close to scale invariant, it is justified to expand the distribution with respect to $\xi(r_{ij})=\Sigma(r_{ij})/\Sigma(0)$ with $i\neq j$.
\end{enumerate}

\subsection{Gauss-Hermite expansion}
\label{subapp:gauss_hermite}

In this subsection we derive the Gauss-Hermite expansion of the Higgs field probability distribution function(PDF). Consider the field at $m$ different positions, $\chi_i=\chi(\mathbf{x}_i),~i=1,\cdots,m$, with mean values $\bar{\chi}_i$. For convenience, in the following we consider the PDF of the random variables $\delta\chi_i=\chi_i-\bar{\chi}$, which can be expressed as
\begin{equation}
    p(\{\delta\chi_i\})=\int\frac{dk_1}{2\pi}\cdots\frac{dk_m}{2\pi}e^{ik_j\delta\chi_j}\phi(\{k_i\})~,
\end{equation}
where $\phi(\{k_i\})$ is the characteristic function of $\delta\chi_i$,
\begin{equation}
    \phi(\{k_i\})=\int d\delta\chi_1\cdots d\delta\chi_m e^{-ik_j\delta\chi_j}p(\{\delta\chi_i\})=E[e^{-ik_j\delta\chi_j}]~.
\end{equation}
The normalization of PDF ensures that $\phi(0)=1$. One can easily obtain the PDF of $\chi_i$ from $p(\{\delta\chi_i\})$ by a simple shift.

The $n$-point correlation function, which is the $n$-th order moment of the random variables $\{\delta\chi_i\}$, can be generated by $\phi(\{k_i\})$,
\begin{equation}
\langle\delta\chi_{j_1}\cdots\delta\chi_{j_n}\rangle=\mu_n^{j_1\cdots j_n}=i^n\left.\frac{\partial^n\phi(\{k_j\})}{\partial k_{j_1}\cdots\partial k_{j_n}}\right|_{k_j=0}~.
\end{equation}

Similarly, we can define the cumulants of $\{\delta\chi_i\}$, which are just the connected correlation functions,
\begin{equation}
\langle\delta\chi_{j_1}\cdots\delta\chi_{j_n}\rangle_c=c_n^{j_1\cdots j_n}=i^n\left.\frac{\partial^n\ln\phi(\{k_j\})}{\partial k_{j_1}\cdots\partial k_{j_n}}\right|_{k_j=0}~.
\end{equation}
It is easy to see that
\begin{equation}
\begin{aligned}
&c_1^j=\mu_1^j=0,\\
&c_2^{ij}=\mu_2^{ij}=\Sigma(r_{ij}),\\
&c_3^{ijk}=\mu_3^{ijk},\\
&c_4^{ijkl}=\mu_4^{ijkl}-c_2^{ij}c_2^{kl}-c_2^{ik}c_2^{jl}-c_2^{il}c_2^{jk}\\
&\cdots
\end{aligned}
\end{equation}

Therefore we can expand $\ln\phi$ as
\begin{equation}
    \ln\phi(\{k_j\})=\sum_{n=1}^{\infty}\frac{(-i)^n}{n!}c_n^{j_1\cdots j_n}k_{j_1}\cdots k_{j_n}.
\end{equation}
Since the Higgs field perturbation is close to Gaussian during inflation, we expect this expansion to converge well, and we can truncate this expansion based on the required accuracy. Then the PDF can be expanded as
\begin{equation}
\label{eq:pdf_expand1}
\begin{aligned}
    p(\{\delta\chi_i\})&=\int\frac{dk_1}{2\pi}\cdots\frac{dk_m}{2\pi}e^{ik_j\delta\chi_j}\exp\left[\sum_{n=1}^\infty\frac{(-i)^n}{n!}c_n^{j_1\cdots j_n}k_{j_1}\cdots k_{j_n}\right]\\
    &=\int \frac{dk_1}{2\pi}\cdots\frac{dk_m}{2\pi}e^{ik_j\delta\chi_j}e^{-\frac{1}{2}c_2^{ij}k_i k_j}\exp\left[\sum_{n=3}^\infty\frac{(-i)^n}{n!}c_n^{j_1\cdots j_n}k_{j_1}\cdots k_{j_n}\right]~.
\end{aligned}
\end{equation}
Using Gaussian Integral, we have
\begin{equation}
\label{eq:gauss_integral}
\begin{aligned}
&\int\frac{dk_1}{2\pi}\cdots\frac{dk_m}{2\pi}e^{ik_j\delta\chi_j}e^{-\frac{1}{2}c_2^{ij}k_i k_j}=\frac{1}{(2\pi)^{-m/2}\sqrt{\det c_2}}e^{-\frac{1}{2}c^{-1}_{2ij} \delta\chi_i \delta\chi_j}=p_G(\{\delta\chi_j\}),\\
&\int\frac{dk_1}{2\pi}\cdots\frac{dk_m}{2\pi}e^{ik_j\delta\chi_j}e^{-\frac{1}{2}c_2^{ij}k_i k_j}k_{j_1}\cdots k_{j_n} =(-i)^n\frac{\partial^n}{\partial\delta\chi_{j_1}\cdots\partial\delta\chi_{j_n}}p_G(\{\delta\chi_j\})~.
\end{aligned}
\end{equation}
To further simplify the above expression, we can decompose $c_2$ as $c_2=AA^T$, which can obtain by diagonalization $c_2=QDQ^T$ and defining $A=QD^{1/2}$. Defining $z_i=A^{-1}_{ij}\delta\chi_j$, Eq.~\eqref{eq:gauss_integral} can be expressed as
\begin{equation}
\begin{aligned}
p_G(\{\delta\chi_j\})&=\frac{1}{(2\pi)^{-m/2}\sqrt{\det c_2}}e^{-\frac{1}{2}z_i z_i},\\
(-i)^n\frac{\partial^n}{\partial\delta\chi_{j_1}\cdots\partial\delta\chi_{j_n}}p_G(\{\delta\chi_j\})&=i^np_G(\{\delta\chi_j\})A^{-1}_{i_1j_1}\cdots A^{-1}_{i_nj_n}H_{j_1\cdots j_n}~,
\end{aligned}
\end{equation}
where $H_{j_1\cdots j_n}$ is the generalized Rodrigues' formula,
\begin{equation}
    H_{j_1\cdots j_n}=(-1)^ne^{\frac{z_iz_i}{2}}\frac{\partial^n}{\partial z_{j_1}\cdots\partial z_{j_n}}e^{-\frac{z_jz_j}{2}}~.
\end{equation}
Now, by expanding the $\exp[\cdots]$ in Eq.~\eqref{eq:pdf_expand1} and utilising the above expressions, we obtain the Gauss-Hermite expansion of the PDF,
\begin{equation}
\label{eq:gauss_hermite_expansion}
    p(\{\delta\chi_i\})\approx p_G(\{\delta\chi_i\})\left(1+\sum_{n=3}^{\infty}\frac{b_n^{i_1\cdots i_n}}{n!}A^{-1}_{i_1j_1}\cdots A^{-1}_{i_nj_n}H_{j_1\cdots j_n} \right)~.
\end{equation}
Here the coefficients $b_n^{i_1\cdots i_n}$ are combinations of $c_r^{i_1\cdots i_r} $ with $r\leq n$, and for $n\leq5$ we simply have $b_n^{i_1\cdots i_n}=c_n^{i_1\cdots i_n}$. 

Notice that we are only interested in the correlation functions of field fluctuations at different spatial points. This simplifies the following derivations considerably.

\subsection{Non-perturbative $\delta N$ coefficients}
\label{subapp:nonpert_coeff}

Now we consider the expansion of the $m$-dimensional Gaussian PDF $p_G(\{\delta\chi_i\})$ around $\xi_{ij}=\Sigma(r_{ij})/\Sigma(0)$. Taking the logarithm, we have
\begin{equation}
\ln p_G=-\frac{m}{2}\ln(2\pi)-\frac{1}{2}\ln\det c_2-\frac{1}{2}\delta\chi_ic^{-1}_{2ij} \delta\chi_j~.
\end{equation}
Since $\xi_{ij}\ll 1$ for $i\neq j$, we can expand the covariance matrix as
\begin{equation}
    c_2^{ij}=\Sigma_{ij}=\Sigma(0)(\delta_{ij}+\epsilon_{ij}),\quad\epsilon_{ij}=\left\{\begin{aligned}
        &0,&i=j,\\&\xi_{ij},&i\neq j.
    \end{aligned}\right.
\end{equation}
Expanding up to the first order of $\epsilon_{ij}$, the logarithm of $p_G(\{\delta\chi_i\})$ can be approximated as
\begin{equation}
\label{eq:log_pdf_expand}
\ln p_G\approx-\frac{m}{2}\ln(2\pi)-\frac{m}{2}\ln\Sigma(0)-\frac{1}{2\Sigma(0)}\delta\chi_i\delta\chi_j(\delta_{ij}-\epsilon_{ij}).
\end{equation}
Therefore we have
\begin{equation}
p_G(\{\delta\chi_i\})\approx p_G(\delta\chi_1)\cdots p_G(\delta\chi_m)\exp\left(\frac{\delta\chi_i\delta\chi_j\epsilon_{ij}}{2\Sigma(0)}\right)~.
\end{equation}
Here $p_G(\delta\chi)$ means the Gaussian distribution around $\delta\chi=0$.

Next we consider the higher order terms in the Gauss-Hermite expansion. Considering $i_1\neq i_2\cdots\neq i_n$, and retaining only the $0$-th order of $\epsilon_{ij}$, we obtain
\begin{equation}
A^{-1}_{i_1j_1}\cdots A^{-1}_{i_nj_n}H_{j_1\cdots j_n}\approx\Sigma(0)^{-n}\delta\chi_{i_1}\cdots\delta\chi_{i_n}~.
\end{equation}
To summarize, after employing the Gauss-Hermite expansion and the expansion of $\xi_{ij}$, the PDF of $\chi_i$ can be approximated as
\begin{equation}
\label{eq:pdf_expand_final}
\begin{aligned}
p(\{\chi_i\})\approx p_G(\chi_1)\cdots p_G(\chi_m)\left(1+\sum_{n=3}^{\infty}\frac{b_n^{i_1\cdots i_n}}{n!}\Sigma(0)^{-n}\delta\chi_{i_1}\cdots\delta\chi_{i_n}\right)\sum_{k=0}^{\infty} \frac{1}{k!}\left(\frac{\delta\chi_i\delta\chi_j\xi_{ij}}{2\Sigma(0)}\right)^k~.   
\end{aligned}
\end{equation}
Here $p_G(\chi)$ means the Gaussian distribution around $\chi=\bar{\chi}$.

Now we can evaluate Eq.~\eqref{eq:exact_cor} using the above approximations. Notice that the integral $\int d\chi_i(N_i-\bar{N})$ equals $0$, therefore after the integral only the terms in the expansion of Eq.~\eqref{eq:pdf_expand_final} with $(\delta\chi)^{n\geq m}$ do not vanish. In the following we derive the leading terms of two-, three-, and four-point functions of $\zeta$ under the above approximations.

\textbf{Two-point correlation function:}
\begin{equation}
\begin{aligned}
\langle\zeta_1\zeta_2\rangle&=\int\mathrm{d}\chi_{1}\mathrm{d}\chi_{2}p(\chi_{1},\chi_{2})(N_{1}-\bar{N})(N_{2}-\bar{N})\\
&\approx \int\mathrm{d}\chi_{1}\mathrm{d}\chi_{2}(N_{1}-\bar{N})(N_{2}-\bar{N})p_G(\chi_1)p_G(\chi_2)\frac{\delta\chi_1\delta\chi_2\xi_{12}}{\Sigma(0)}\\
&\approx\tilde{N}^{\prime2}\Sigma(r_{12})~,
\end{aligned}
\end{equation}
where $\tilde{N}^\prime$ is defined as
\begin{equation}
	\tilde{N}^\prime=\Sigma(0)^{-1}\int\mathrm{d}\chi p_\mathrm{G}(\chi)N\delta\chi~.
\end{equation}

\textbf{Three-point correlation function:}
\begin{equation}
\begin{aligned}
\langle\zeta_1\zeta_2\zeta_3\rangle&=\int\mathrm{d}\chi_{1}\mathrm{d}\chi_{2}\mathrm{d}\chi_{3}p(\chi_{1},\chi_{2},\chi_{3})(N_{1}-\bar{N})(N_{2}-\bar{N})(N_{3}-\bar{N})\\
&\approx \int\mathrm{d}\chi_{1}\mathrm{d}\chi_{2}\mathrm{d}\chi_{3}(N_{1}-\bar{N})(N_{2}-\bar{N})(N_{3}-\bar{N})p_G(\chi_1)p_G(\chi_2)p_G(\chi_3)\\&\quad\quad\times\left(\frac{c_3^{ijk}}{6\Sigma(0)^3}\delta\chi_i\delta\chi_j\delta\chi_k+\frac{\delta\chi_i\delta\chi_j\delta\chi_k\delta\chi_l\xi_{ij}\xi_{kl}}{8\Sigma(0)^2}\right)\\
&\approx\tilde{N}^{\prime3}c_3^{123}+\tilde{N}^{\prime\prime}\tilde{N}^{\prime2}\left[\Sigma(r_{12})\Sigma(r_{13})+\Sigma(r_{12})\Sigma(r_{23})+\Sigma(r_{13})\Sigma(r_{23})\right]~,
\end{aligned}
\end{equation}
where $c_3^{123}=\langle\delta\chi_1\delta\chi_2\delta\chi_3\rangle$ and $\tilde{N}^{\prime\prime}$ is defined as
\begin{equation}
\tilde{N}^{\prime\prime}=\Sigma(0)^{-2}\int\mathrm{d}\chi p_\mathrm{G}(\chi)(N-\bar{N})\delta\chi^2~.
\end{equation}
The first term of the $3$-point function of $\zeta$ arises from the interaction of the field $\chi$ during inflation, while the second term results from the nonlinear evolution of the superhorizon perturbations.

\textbf{Four-point correlation function:}
\begin{equation}
\begin{aligned}
\langle\zeta_1\zeta_2\zeta_3\zeta_4\rangle&=\int\mathrm{d}\chi_{1}\cdots\mathrm{d}\chi_{4}p(\chi_{1},\cdots,\chi_{4})(N_{1}-\bar{N})\cdots(N_{4}-\bar{N})\\
&\approx \int\mathrm{d}\chi_{1}\cdots\mathrm{d}\chi_{4}(N_{1}-\bar{N})\cdots(N_{4}-\bar{N})p_G(\chi_1)\cdots p_G(\chi_4)\\&\quad\quad\times\left(\frac{c_4^{ijkl}}{24\Sigma(0)^4}\delta\chi_i\delta\chi_j\delta\chi_k\delta\chi_l+\frac{\delta\chi_i\delta\chi_j\delta\chi_k\delta\chi_l\xi_{ij}\xi_{kl}}{8\Sigma(0)^2}\right.\\&\left.\quad\quad\quad+\frac{c_3^{ijk}}{6\Sigma(0)^3}\delta\chi_i\delta\chi_j\delta\chi_k\frac{\delta\chi_l\delta\chi_m\xi_{lm}}{2\Sigma(0)}+\frac{\delta\chi_i\delta\chi_j\delta\chi_k\delta\chi_l\delta\chi_m\delta\chi_n\xi_{ij}\xi_{kl}\xi_{mn}}{48\Sigma(0)^3}\right)\\
&\approx\tilde{N}^{\prime4}c_4^{1234}+\tilde{N}^{\prime4}\left[\Sigma(r_{12})\Sigma(r_{34})+2\text{perms}\right] +\tilde{N}^{\prime\prime}\tilde{N}^{\prime3}\left[c_3^{123}\Sigma(r_{14})+11\text{perms} \right]\\
&\quad+\tilde{N}^{\prime\prime\prime}\tilde{N}^{\prime3}\left[\Sigma(r_{12})\Sigma(r_{13})\Sigma(r_{14})+3\text{perms}\right]\\
&\quad+\tilde{N}^{\prime\prime2}\tilde{N}^{\prime2}\left[\Sigma(r_{12})\Sigma(r_{13})\Sigma(r_{24})+11\text{perms}\right]~.
\end{aligned}
\end{equation}
where $c_4^{1234}=\langle\delta\chi_1\delta\chi_2\delta\chi_3\delta\chi_4\rangle_c$ and $\tilde{N}^{\prime\prime\prime}$ is defined as
\begin{equation}
\tilde{N}^{\prime\prime\prime}=\Sigma(0)^{-3}\int\mathrm{d}\chi p_\mathrm{G}(\chi)(N-\bar{N})\delta\chi^3~.
\end{equation}
The first term of the $4$-point function of $\zeta$ arises from the self interaction of the field $\chi$ during inflation, while the second term is just the connected part of the $4$-point function. The last 3 terms result from the nonlinear evolution of the superhorizon perturbations.

Following the previous procedure, the correlation functions of $\zeta$ with $n>4$ can also be effectively evaluated using the non-perturbative $\delta N$ method. We find that the expressions of the $2$-, $3$-, and $4$-point functions of $\zeta$ evaluated in the non-perturbative $\delta N$ method is rather analogous to the expressions in the naive $\delta N$ formalism with Taylor expansion. The only difference is that the Taylor coefficients are replaced by the so-called non-perturbative $\delta N$ coefficients $\tilde{N}^\prime$, $\tilde{N}^{\prime\prime},\cdots$. Therefore the formula of $\delta N$ method in $k$-space should also hold in the non-perturbative method, as long as the Taylor coefficients are replaced. For example, the power spectrum and bispectrum of the curvature perturbation can be expressed as 
\begin{equation}
\begin{aligned}
P_{\zeta}(k)&\approx\tilde{N}^{\prime2}P_\chi(k),\\	B_\zeta(k_1,k_2,k_3)&\approx\tilde{N}^{\prime3}B_\chi(k_{1},k_{2},k_{3})+\tilde{N}^{\prime2}\tilde{N}^{\prime\prime}(P_\chi(k_{1})P_\chi(k_{2})+\mathrm{cyclic})~.
\end{aligned}
\end{equation}

\acknowledgments
We acknowledge Misao Sasaki, Jinn-Ouk Gong and Shi Pi for helpful discussion. This work was supported by National Key R\&D Program of China under grant No. 2023YFA1606100, by the National Natural Science Foundation of China (NSFC) under grant Nos.12435005 and 12335005.
C.\,H.\ acknowledges supports from the Sun Yat-Sen University Science Foundation, and the Key Laboratory of Particle Astrophysics and Cosmology (MOE) of Shanghai Jiao Tong University.


\bibliographystyle{JHEP}
\bibliography{biblio.bib}

\providecommand{\href}[2]{#2}\begingroup\raggedright\begin{thebibliography}{10}

\bibitem{Guth:1982ec}
A.H.~Guth and S.Y.~Pi, \emph{{Fluctuations in the New Inflationary Universe}},
  \href{https://doi.org/10.1103/PhysRevLett.49.1110}{\emph{Phys. Rev. Lett.}
  {\bfseries 49} (1982) 1110}.

\bibitem{Hawking:1982cz}
S.W.~Hawking, \emph{{The Development of Irregularities in a Single Bubble
  Inflationary Universe}},
  \href{https://doi.org/10.1016/0370-2693(82)90373-2}{\emph{Phys. Lett. B}
  {\bfseries 115} (1982) 295}.

\bibitem{Bardeen:1983qw}
J.M.~Bardeen, P.J.~Steinhardt and M.S.~Turner, \emph{{Spontaneous Creation of
  Almost Scale - Free Density Perturbations in an Inflationary Universe}},
  \href{https://doi.org/10.1103/PhysRevD.28.679}{\emph{Phys. Rev. D} {\bfseries
  28} (1983) 679}.

\bibitem{Lyth:1984gv}
D.H.~Lyth, \emph{{Large Scale Energy Density Perturbations and Inflation}},
  \href{https://doi.org/10.1103/PhysRevD.31.1792}{\emph{Phys. Rev. D}
  {\bfseries 31} (1985) 1792}.

\bibitem{Mukhanov:1981xt}
V.F.~Mukhanov and G.V.~Chibisov, \emph{{Quantum Fluctuations and a Nonsingular
  Universe}}, {\emph{JETP Lett.} {\bfseries 33} (1981) 532}.

\bibitem{Starobinsky:1982ee}
A.A.~Starobinsky, \emph{{Dynamics of Phase Transition in the New Inflationary
  Universe Scenario and Generation of Perturbations}},
  \href{https://doi.org/10.1016/0370-2693(82)90541-X}{\emph{Phys. Lett. B}
  {\bfseries 117} (1982) 175}.

\bibitem{Dvali:2003em}
G.~Dvali, A.~Gruzinov and M.~Zaldarriaga, \emph{{A new mechanism for generating
  density perturbations from inflation}},
  \href{https://doi.org/10.1103/PhysRevD.69.023505}{\emph{Phys. Rev. D}
  {\bfseries 69} (2004) 023505}
  [\href{https://arxiv.org/abs/astro-ph/0303591}{{\ttfamily
  astro-ph/0303591}}].

\bibitem{Kofman:2003nx}
L.~Kofman, \emph{{Probing string theory with modulated cosmological
  fluctuations}},  \href{https://arxiv.org/abs/astro-ph/0303614}{{\ttfamily
  astro-ph/0303614}}.

\bibitem{Suyama:2007bg}
T.~Suyama and M.~Yamaguchi, \emph{{Non-Gaussianity in the modulated reheating
  scenario}}, \href{https://doi.org/10.1103/PhysRevD.77.023505}{\emph{Phys.
  Rev. D} {\bfseries 77} (2008) 023505}
  [\href{https://arxiv.org/abs/0709.2545}{{\ttfamily 0709.2545}}].

\bibitem{Ichikawa:2008ne}
K.~Ichikawa, T.~Suyama, T.~Takahashi and M.~Yamaguchi, \emph{{Primordial
  Curvature Fluctuation and Its Non-Gaussianity in Models with Modulated
  Reheating}}, \href{https://doi.org/10.1103/PhysRevD.78.063545}{\emph{Phys.
  Rev. D} {\bfseries 78} (2008) 063545}
  [\href{https://arxiv.org/abs/0807.3988}{{\ttfamily 0807.3988}}].

\bibitem{Starobinsky:1986fx}
A.A.~Starobinsky, \emph{{STOCHASTIC DE SITTER (INFLATIONARY) STAGE IN THE EARLY
  UNIVERSE}}, \href{https://doi.org/10.1007/3-540-16452-9_6}{\emph{Lect. Notes
  Phys.} {\bfseries 246} (1986) 107}.

\bibitem{Starobinsky:1994bd}
A.A.~Starobinsky and J.~Yokoyama, \emph{{Equilibrium state of a selfinteracting
  scalar field in the De Sitter background}},
  \href{https://doi.org/10.1103/PhysRevD.50.6357}{\emph{Phys. Rev. D}
  {\bfseries 50} (1994) 6357}
  [\href{https://arxiv.org/abs/astro-ph/9407016}{{\ttfamily
  astro-ph/9407016}}].

\bibitem{Markkanen:2019kpv}
T.~Markkanen, A.~Rajantie, S.~Stopyra and T.~Tenkanen, \emph{{Scalar
  correlation functions in de Sitter space from the stochastic spectral
  expansion}}, \href{https://doi.org/10.1088/1475-7516/2019/08/001}{\emph{JCAP}
  {\bfseries 08} (2019) 001}
  [\href{https://arxiv.org/abs/1904.11917}{{\ttfamily 1904.11917}}].

\bibitem{Choi:2012cp}
K.-Y.~Choi and Q.-G.~Huang, \emph{{Can the standard model Higgs boson seed the
  formation of structures in our Universe?}},
  \href{https://doi.org/10.1103/PhysRevD.87.043501}{\emph{Phys. Rev. D}
  {\bfseries 87} (2013) 043501}
  [\href{https://arxiv.org/abs/1209.2277}{{\ttfamily 1209.2277}}].

\bibitem{DeSimone:2012gq}
A.~De~Simone, H.~Perrier and A.~Riotto, \emph{{Non-Gaussianities from the
  Standard Model Higgs}},
  \href{https://doi.org/10.1088/1475-7516/2013/01/037}{\emph{JCAP} {\bfseries
  01} (2013) 037} [\href{https://arxiv.org/abs/1210.6618}{{\ttfamily
  1210.6618}}].

\bibitem{Cai:2013caa}
Y.-F.~Cai, Y.-C.~Chang, P.~Chen, D.A.~Easson and T.~Qiu, \emph{{Planck
  constraints on Higgs modulated reheating of renormalization group improved
  inflation}}, \href{https://doi.org/10.1103/PhysRevD.88.083508}{\emph{Phys.
  Rev. D} {\bfseries 88} (2013) 083508}
  [\href{https://arxiv.org/abs/1304.6938}{{\ttfamily 1304.6938}}].

\bibitem{Karam:2021qgn}
A.~Karam, T.~Markkanen, L.~Marzola, S.~Nurmi, M.~Raidal and A.~Rajantie,
  \emph{{Higgs-like spectator field as the origin of structure}},
  \href{https://doi.org/10.1140/epjc/s10052-021-09417-w}{\emph{Eur. Phys. J. C}
  {\bfseries 81} (2021) 620}
  [\href{https://arxiv.org/abs/2103.02569}{{\ttfamily 2103.02569}}].

\bibitem{Litsa:2020mvj}
A.~Litsa, K.~Freese, E.I.~Sfakianakis, P.~Stengel and L.~Visinelli,
  \emph{{Primordial non-Gaussianity from the effects of the Standard Model
  Higgs during reheating after inflation}},
  \href{https://doi.org/10.1088/1475-7516/2023/03/033}{\emph{JCAP} {\bfseries
  03} (2023) 033} [\href{https://arxiv.org/abs/2011.11649}{{\ttfamily
  2011.11649}}].

\bibitem{Litsa:2020rsm}
A.~Litsa, K.~Freese, E.I.~Sfakianakis, P.~Stengel and L.~Visinelli,
  \emph{{Large density perturbations from reheating to standard model particles
  due to the dynamics of the Higgs boson during inflation}},
  \href{https://doi.org/10.1103/PhysRevD.104.123546}{\emph{Phys. Rev. D}
  {\bfseries 104} (2021) 123546}
  [\href{https://arxiv.org/abs/2009.14218}{{\ttfamily 2009.14218}}].

\bibitem{Lu:2019tjj}
S.~Lu, Y.~Wang and Z.-Z.~Xianyu, \emph{{A Cosmological Higgs Collider}},
  \href{https://doi.org/10.1007/JHEP02(2020)011}{\emph{JHEP} {\bfseries 02}
  (2020) 011} [\href{https://arxiv.org/abs/1907.07390}{{\ttfamily
  1907.07390}}].

\bibitem{You:2024hit}
J.~You, L.~Song, H.-J.~He and C.~Han, \emph{{Cosmological non-Gaussianity from
  the neutrino seesaw mechanism}},
  \href{https://doi.org/10.1103/lhmk-ycfr}{\emph{Phys. Rev. D} {\bfseries 112}
  (2025) 083555} [\href{https://arxiv.org/abs/2412.16033}{{\ttfamily
  2412.16033}}].

\bibitem{Han:2024qbw}
C.~Han, H.-J.~He, L.~Song and J.~You, \emph{{Cosmological signatures of
  neutrino seesaw mechanism}},
  \href{https://doi.org/10.1103/b7hv-3h2p}{\emph{Phys. Rev. D} {\bfseries 112}
  (2025) L081309} [\href{https://arxiv.org/abs/2412.21045}{{\ttfamily
  2412.21045}}].

\bibitem{Starobinsky:1985ibc}
A.A.~Starobinsky, \emph{{Multicomponent de Sitter (Inflationary) Stages and the
  Generation of Perturbations}}, {\emph{JETP Lett.} {\bfseries 42} (1985) 152}.

\bibitem{Salopek:1990jq}
D.S.~Salopek and J.R.~Bond, \emph{{Nonlinear evolution of long wavelength
  metric fluctuations in inflationary models}},
  \href{https://doi.org/10.1103/PhysRevD.42.3936}{\emph{Phys. Rev. D}
  {\bfseries 42} (1990) 3936}.

\bibitem{Comer:1994np}
G.L.~Comer, N.~Deruelle, D.~Langlois and J.~Parry, \emph{{Growth or decay of
  cosmological inhomogeneities as a function of their equation of state}},
  \href{https://doi.org/10.1103/PhysRevD.49.2759}{\emph{Phys. Rev. D}
  {\bfseries 49} (1994) 2759}.

\bibitem{Sasaki:1995aw}
M.~Sasaki and E.D.~Stewart, \emph{{A General analytic formula for the spectral
  index of the density perturbations produced during inflation}},
  \href{https://doi.org/10.1143/PTP.95.71}{\emph{Prog. Theor. Phys.} {\bfseries
  95} (1996) 71} [\href{https://arxiv.org/abs/astro-ph/9507001}{{\ttfamily
  astro-ph/9507001}}].

\bibitem{Sasaki:1998ug}
M.~Sasaki and T.~Tanaka, \emph{{Superhorizon scale dynamics of multiscalar
  inflation}}, \href{https://doi.org/10.1143/PTP.99.763}{\emph{Prog. Theor.
  Phys.} {\bfseries 99} (1998) 763}
  [\href{https://arxiv.org/abs/gr-qc/9801017}{{\ttfamily gr-qc/9801017}}].

\bibitem{Wands:2000dp}
D.~Wands, K.A.~Malik, D.H.~Lyth and A.R.~Liddle, \emph{{A New approach to the
  evolution of cosmological perturbations on large scales}},
  \href{https://doi.org/10.1103/PhysRevD.62.043527}{\emph{Phys. Rev. D}
  {\bfseries 62} (2000) 043527}
  [\href{https://arxiv.org/abs/astro-ph/0003278}{{\ttfamily
  astro-ph/0003278}}].

\bibitem{Lyth:2003im}
D.H.~Lyth and D.~Wands, \emph{{Conserved cosmological perturbations}},
  \href{https://doi.org/10.1103/PhysRevD.68.103515}{\emph{Phys. Rev. D}
  {\bfseries 68} (2003) 103515}
  [\href{https://arxiv.org/abs/astro-ph/0306498}{{\ttfamily
  astro-ph/0306498}}].

\bibitem{Rigopoulos:2003ak}
G.I.~Rigopoulos and E.P.S.~Shellard, \emph{{The separate universe approach and
  the evolution of nonlinear superhorizon cosmological perturbations}},
  \href{https://doi.org/10.1103/PhysRevD.68.123518}{\emph{Phys. Rev. D}
  {\bfseries 68} (2003) 123518}
  [\href{https://arxiv.org/abs/astro-ph/0306620}{{\ttfamily
  astro-ph/0306620}}].

\bibitem{Lyth:2004gb}
D.H.~Lyth, K.A.~Malik and M.~Sasaki, \emph{{A General proof of the conservation
  of the curvature perturbation}},
  \href{https://doi.org/10.1088/1475-7516/2005/05/004}{\emph{JCAP} {\bfseries
  05} (2005) 004} [\href{https://arxiv.org/abs/astro-ph/0411220}{{\ttfamily
  astro-ph/0411220}}].

\bibitem{Lyth:2005fi}
D.H.~Lyth and Y.~Rodriguez, \emph{{The Inflationary prediction for primordial
  non-Gaussianity}},
  \href{https://doi.org/10.1103/PhysRevLett.95.121302}{\emph{Phys. Rev. Lett.}
  {\bfseries 95} (2005) 121302}
  [\href{https://arxiv.org/abs/astro-ph/0504045}{{\ttfamily
  astro-ph/0504045}}].

\bibitem{Planck:2018nkj}
{\scshape Planck} collaboration, \emph{{Planck 2018 results. I. Overview and
  the cosmological legacy of Planck}},
  \href{https://doi.org/10.1051/0004-6361/201833880}{\emph{Astron. Astrophys.}
  {\bfseries 641} (2020) A1}
  [\href{https://arxiv.org/abs/1807.06205}{{\ttfamily 1807.06205}}].

\bibitem{Planck:2018jri}
{\scshape Planck} collaboration, \emph{{Planck 2018 results. X. Constraints on
  inflation}}, \href{https://doi.org/10.1051/0004-6361/201833887}{\emph{Astron.
  Astrophys.} {\bfseries 641} (2020) A10}
  [\href{https://arxiv.org/abs/1807.06211}{{\ttfamily 1807.06211}}].

\bibitem{Komatsu:2001rj}
E.~Komatsu and D.N.~Spergel, \emph{{Acoustic signatures in the primary
  microwave background bispectrum}},
  \href{https://doi.org/10.1103/PhysRevD.63.063002}{\emph{Phys. Rev. D}
  {\bfseries 63} (2001) 063002}
  [\href{https://arxiv.org/abs/astro-ph/0005036}{{\ttfamily
  astro-ph/0005036}}].

\bibitem{Maldacena:2002vr}
J.M.~Maldacena, \emph{{Non-Gaussian features of primordial fluctuations in
  single field inflationary models}},
  \href{https://doi.org/10.1088/1126-6708/2003/05/013}{\emph{JHEP} {\bfseries
  05} (2003) 013} [\href{https://arxiv.org/abs/astro-ph/0210603}{{\ttfamily
  astro-ph/0210603}}].

\bibitem{Wands:2010af}
D.~Wands, \emph{{Local non-Gaussianity from inflation}},
  \href{https://doi.org/10.1088/0264-9381/27/12/124002}{\emph{Class. Quant.
  Grav.} {\bfseries 27} (2010) 124002}
  [\href{https://arxiv.org/abs/1004.0818}{{\ttfamily 1004.0818}}].

\bibitem{Suyama:2013dqa}
T.~Suyama and S.~Yokoyama, \emph{{Statistics of general functions of a Gaussian
  field -application to non-Gaussianity from preheating-}},
  \href{https://doi.org/10.1088/1475-7516/2013/06/018}{\emph{JCAP} {\bfseries
  06} (2013) 018} [\href{https://arxiv.org/abs/1303.1254}{{\ttfamily
  1303.1254}}].

\bibitem{Imrith:2018uyk}
S.V.~Imrith, D.J.~Mulryne and A.~Rajantie, \emph{{Nonperturbative $\delta N$
  formalism}}, \href{https://doi.org/10.1103/PhysRevD.98.043513}{\emph{Phys.
  Rev. D} {\bfseries 98} (2018) 043513}
  [\href{https://arxiv.org/abs/1801.02600}{{\ttfamily 1801.02600}}].

\bibitem{Planck:2019kim}
{\scshape Planck} collaboration, \emph{{Planck 2018 results. IX. Constraints on
  primordial non-Gaussianity}},
  \href{https://doi.org/10.1051/0004-6361/201935891}{\emph{Astron. Astrophys.}
  {\bfseries 641} (2020) A9}
  [\href{https://arxiv.org/abs/1905.05697}{{\ttfamily 1905.05697}}].

\bibitem{Weinberg:2005vy}
S.~Weinberg, \emph{{Quantum contributions to cosmological correlations}},
  \href{https://doi.org/10.1103/PhysRevD.72.043514}{\emph{Phys. Rev. D}
  {\bfseries 72} (2005) 043514}
  [\href{https://arxiv.org/abs/hep-th/0506236}{{\ttfamily hep-th/0506236}}].

\bibitem{Weinberg:2006ac}
S.~Weinberg, \emph{{Quantum contributions to cosmological correlations. II. Can
  these corrections become large?}},
  \href{https://doi.org/10.1103/PhysRevD.74.023508}{\emph{Phys. Rev. D}
  {\bfseries 74} (2006) 023508}
  [\href{https://arxiv.org/abs/hep-th/0605244}{{\ttfamily hep-th/0605244}}].

\bibitem{Chen:2017ryl}
X.~Chen, Y.~Wang and Z.-Z.~Xianyu, \emph{{Schwinger-Keldysh Diagrammatics for
  Primordial Perturbations}},
  \href{https://doi.org/10.1088/1475-7516/2017/12/006}{\emph{JCAP} {\bfseries
  12} (2017) 006} [\href{https://arxiv.org/abs/1703.10166}{{\ttfamily
  1703.10166}}].

\bibitem{Mulryne:2009kh}
D.J.~Mulryne, D.~Seery and D.~Wesley, \emph{{Moment transport equations for
  non-Gaussianity}},
  \href{https://doi.org/10.1088/1475-7516/2010/01/024}{\emph{JCAP} {\bfseries
  01} (2010) 024} [\href{https://arxiv.org/abs/0909.2256}{{\ttfamily
  0909.2256}}].

\bibitem{Mulryne:2010rp}
D.J.~Mulryne, D.~Seery and D.~Wesley, \emph{{Moment transport equations for the
  primordial curvature perturbation}},
  \href{https://doi.org/10.1088/1475-7516/2011/04/030}{\emph{JCAP} {\bfseries
  04} (2011) 030} [\href{https://arxiv.org/abs/1008.3159}{{\ttfamily
  1008.3159}}].

\end{thebibliography}\endgroup

\end{document}